\documentclass[12pt,preprint]{aastex}

\newcommand{\ch}{{\it Chandra}}
\newcommand{\xmm}{{\it XMM-Newton}}
\newcommand{\fu}{{\it FUSE }}
\newcommand{\ulg}{ULIRGs}

\newcommand{\ergs} {erg s$^{-1}$}

\newcommand{\afe}{$\alpha/\rm{Fe}$}
\newcommand{\alphe}{$\alpha$-element}
\newcommand{\ha}{$\rm{H}\-\alpha$}
\newcommand{\lya}{$\rm{Ly}\-\alpha$}
\newcommand{\ang}{\rm{\AA} }
\newcommand{\vv}{VV 114}

\newcommand{\ovi}{\ion{O}{6}}
\newcommand{\hoo}{Haro\,11}
\newcommand{\ks}{\rm{km\,s^{-1}}}
\newcommand{\kms}{$\rm{km\,s^{-1}}$}

\shorttitle{\fu\,\&\ch\,Observations of Haro 11}
\shortauthors{J Grimes et al.}

\slugcomment{Accepted for Publication in ApJ}

\begin{document}

\title{Feedback in the local LBG Analog Haro 11 as probed by far-UV and X-ray observations}

\author{J. P. Grimes\altaffilmark{1}, T. Heckman\altaffilmark{1}, 
D. Strickland\altaffilmark{1}, W. V. Dixon\altaffilmark{1},
K. Sembach\altaffilmark{2},
R. Overzier\altaffilmark{1}, C. Hoopes\altaffilmark{1},
A. Aloisi\altaffilmark{2}, and A. Ptak\altaffilmark{1}}
\altaffiltext{1}{Center for Astrophysical Sciences, Johns Hopkins University,
    3400 N. Charles St, Baltimore, MD 21218; jgrimes@pha.jhu.edu, heckman@pha.jhu.edu,
    dks@pha.jhu.edu,ptak@pha.jhu.edu}
\altaffiltext{2}{Space Telescope Science Institute, 3700 San Martin Drive,
   Baltimore, MD, 21218; sembach@stsci.edu,aloisi@stsci.edu}

\begin{abstract}

We have re-analyzed \fu\, data and obtained new \ch~observations of \hoo, 
a local ($\rm{D_L}$=88 Mpc) UV luminous galaxy.  
\hoo\, has a similar  far-UV luminosity ($10^{10.3}\,\rm{L}_{\odot}$), UV surface 
brightness ($10^{9.4}\,\rm{L_{\odot}\,kpc^{-2}}$), SFR, and metallicity to that observed
in Lyman Break Galaxies (LBGs).  We show that \hoo\, has extended, soft thermal 
($\rm{kT\sim0.68\,keV}$) X-ray emission with a luminosity and size which scales with 
the physical properties (e.g. SFR, stellar mass) of the host galaxy. 
 An enhanced \afe\, ratio of $\sim4$ relative to solar abundance
suggests significant supernovae enrichment.  These results are consistent with
the X-ray emission being produced in a shock between a supernovae driven outflow
and the ambient material.
The FUV spectra show strong absorption lines similar to those observed in LBG spectra.
A blueshifted absorption component is identified as a wind outflowing at $\sim200-280$ \kms.
\ion{O}{6}$\lambda\lambda$1032,1038 emission, the dominant cooling mechanism for coronal 
gas at $\rm{T\sim10^{5.5}K}$ is also observed.  
If associated with the outflow, the luminosity of the \ovi\, emission 
suggests that $\lesssim20\%$ of the total mechanical energy 
from the supernovae and solar winds is being radiated away.  
This implies that 
radiative cooling through \ovi\, is not significantly inhibiting the growth of
the outflowing gas.
In contradiction to the findings of \citet{berg06}, we find 
no convincing evidence of Lyman continuum leakage
in \hoo.  We conclude that the wind has not created a ``tunnel'' allowing the escape of a significant 
fraction of Lyman continuum photons and place a limit on the escape
fraction of $\rm{f_{esc}}\lesssim2\%$.
Overall, both \hoo\, and a previously observed LBG analogue \vv, provide
an invaluable insight into the X-ray and FUV properties of 
high redshift LBGs.

\end{abstract}


\section{Introduction}

Recent observations, headlined by the {\it WMAP} satellite, 
have ushered in a new era of ``precision cosmology'' \citep{benn03}.  
Measurements of the age, geometry, density, and 
composition of the universe have led to
a robust understanding of the evolution of its large scale structure
(dominated by dark matter and dark energy).   
However, significant problems in $\Lambda$CDM 
simulations emerge when examining
small scale structure and regions of high density where the complex range of baryonic physics becomes important \citep{klyp99,rob05,som99}.   
The frontier of cosmology is now in
understanding the physical processes involved in 
the interactions between the intergalactic medium (IGM) and the gas, stars,
and black holes that comprise galaxies.

At the forefront of
this endeavor is understanding the role of feedback with which a
galaxy mediates both its own evolution and that of the
next generation of galaxies.  Supernovae driven galactic scale winds
appear to be an essential piece of galaxy evolution \citep{veill05}.
In galaxies undergoing intense star formation, these winds 
drive gas, energy, and metals out of galaxies and possibly into the IGM \citep{heck98}. 
Although common in the local universe,
these winds were of particular importance in the past,
due to the significantly higher overall star formation rate (SFR) 
in the early universe \citep{bunk04}. 

Such winds at high redshift almost
certainly had a profound
impact on the energetics, ionization, and chemical composition
of the IGM \citep{ag05}.  For example, star-forming galaxies are thought to be
an important source of ionizing photons in the early universe
and could play a vital role in the reionization
of the IGM.  This effect should be
enhanced by winds as they could
clear paths through the neutral
gas and dust surrounding the galaxy,
allowing the ionizing photons to
escape into the IGM \citep{dov00}.  A reionized IGM
strongly impacts galaxy formation
as it suppresses gas cooling and subsequent infall into low mass
halos \citep{bark06}.

Multiwavelength observations are required to understand the complex multiphase
nature of galactic winds \citep{strick04b}. The coronal 
gas ($T \sim 10^5$ to $10^6$ K) and hot X-ray gas ($T \sim 10^6$ to $10^8$ K) in 
galactic winds are particularly important.  These phases
are intimately connected to the mechanical/thermal energy that drives the 
outflows and to the metals the outflows carry.
Far-ultraviolet (FUV) observations 
of the coronal gas provide important insights into the cooling 
and kinematics of the galactic winds \citep{heck01,hoop03}.  X-ray observations 
probe the slowly cooling hot gas that is most likely to escape the
galaxy \citep{strick00}.

The most widely studied population of high redshift star forming galaxies are 
the Lyman Break Galaxies \citep[LBGs,][]{steid99}.  These galaxies can be efficiently detected for 
$2\lesssim z \lesssim 6$ using
the Lyman-break technique, which picks out spectral band dropouts caused by the
912 $\ang$ Lyman continuum discontinuity. They constitute a significant (and possibly dominant) fraction of the population of star forming galaxies during this important cosmic epoch \citep{peac00}.    The LBGs trace the most overdense regions of the universe, which are believed to be the progenitors of present day galaxy groups and clusters \citep{giav02,ouchi04}. They therefore play an important role in understanding the evolution of such clusters.  
Galactic winds appear to be a ubiquitous property of LBGs \citep{shap03}.
In particular, winds from LBGs could be the mechanism that heats
and chemically enriches the Intra-cluster Medium \citep[ICM, ][]{hels00,tam04}.

Unfortunately, our knowledge of the properties of the winds in LBGs is limited to what can be inferred from the interstellar UV absorption lines longward of Lyman $\alpha$. Because of their great distance, X-ray detections are almost impossible using
current X-ray observatories, although stacking techniques have been used to
construct luminosity weighted average X-ray spectra \citep{nan02,lehm05}.  
A more fundamental problem is that the large redshifts of the LBGs mean that soft X-ray observations with {\it Chandra} or \xmm ~are observing at rest-frame energies above a few keV. In local star-forming galaxies the emission in this band is dominated by the population of X-ray binaries, and the thermal emission from the galactic wind is negligible \citep{colb04}. Likewise, observations of the coronal phase gas is very difficult in LBGs. The most accessible probe of such gas is the  \ion{O}{6}~$\lambda\lambda$1032,1038 doublet, which lies deep within the Lyman $\alpha$ forest in the spectra of LBGs. 

It is clear that directly studying gas hotter than $10^5\,$K in high-redshift LBGs will continue to be very difficult or even impossible for the forseeable future.  Therefore, an important step in understanding galactic winds in LBGs would be to identify the best local analogs to LBGs and to then investigate their winds using the full suite of observations that are possible at low redshift.

Recently \citet{heck05} created a catalog of low redshift galaxies 
using {\it GALEX} \citep{mart05} UV observations of galaxies with spectra taken by the Sloan Digital Sky Survey \citep[SDSS, ][]{york00}. This catalog has now been significantly expanded based on additional {\it GALEX} data by Hoopes et al. (2006), who used this matched catalog to select a sample of extraordinarily rare local (z $<$ 0.3) galaxies having the same UV luminosities, sizes, and surface brightnesses as typical LBGs. They then showed that these galaxies have the same star formation rates, galaxy masses, velocity dispersions, and chemical compositions as typical LBGs. 

In the {\it GALEX} sample, \vv~   is the 
closest ($\rm{D_L\sim 86\, Mpc}$) known LBG analog.  We have previously 
analyzed X-ray and FUV observations of
\vv\, \citep{grim06}.  Diffuse thermal X-ray emission
encompassing \vv~was observed by \ch\, and \xmm.  This hot
($\rm{kT}\sim 0.59~\rm{keV}$) gas has an enhanced
\afe~element ratio relative to solar abundances and follows the
same relations as typical starbursts between its properties (luminosity, size, and temperature) and those of the starburst galaxy
(SFR, dust temperature, galaxy mass). 
These results are consistent with the X-ray gas having been produced by shocks driven by a galactic superwind.
The \fu\, observations of \vv~show strong, broad interstellar absorption lines with a pronounced blueshifted component
(similar to what is seen in LBGs). This implies 
an outflow of material moving at $\sim300-400~\ks$ relative to \vv.  The properties 
of the strong \ovi~absorption line are consistent with radiative
cooling at the interface between the hot outrushing gas seen in X-rays and
the cooler material seen in the other outflowing ions in the \fu\, data.
We also showed that the wind in \vv~has {\it not} created a ``tunnel'' that would enable more than a 
small fraction ($<$ few percent) of the ionizing photons from \vv ~to escape into the IGM.

The next closest LBG analog in the {\it GALEX} sample is \hoo.  \hoo\, is particularly interesting,
as \citet{berg06} have claimed the detection of Lyman continuum radiation in the \fu~spectra.  
\citet{berg06} derived an escape fraction of $\rm{f_{esc}}\sim4-10\%$.
This would make it the only known galaxy with Lyman continuum emission in the local universe
and thus an ideal candidate to study this intriguing property of some high redshift LBGs \citep{shap06}.

\begin{figure}
\centering
\leavevmode
\includegraphics[width=3in]{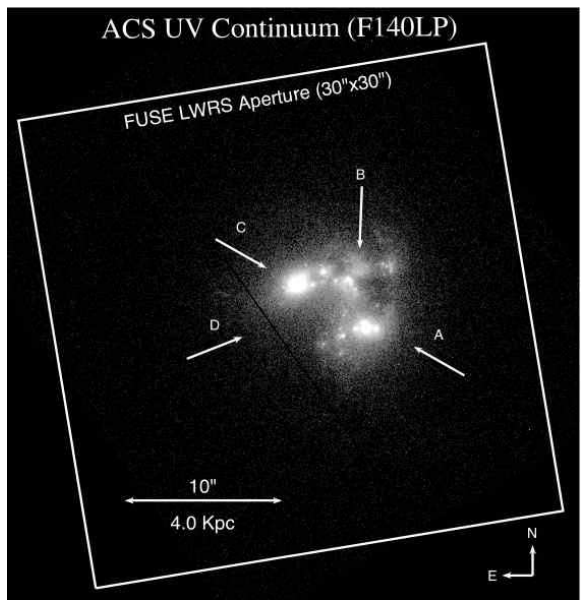}
\caption{
\fu\, LWRS aperture overlaid on an HST/ACS 
FUV (F140LP) continuum image of \hoo.    We have also 
labeled the three knots (A-C) as in \citet{vad93}.  A fourth knot,  identified by \citet{kun03}
is labeled D, although it is not visible in this image.
\label{acs}}
\end{figure}

\hoo, also  known as ESO 350-IG38 ($\alpha_{2000}=00^{\rm h}36^{\rm m}52\fs5$,
$\delta_{2000}=-33\degr33\arcmin19\arcsec$), is a starbursting blue compact galaxy 
located $\sim$88 Mpc.  This distance was derived using $\rm{v_{sys}}=6180\, \ks$
\citep{berg06} and assuming $\rm{H_0=71\,km\,s^{-1}\,Mpc^{-1}}$.
\hoo's far-UV luminosity ($10^{10.3}\,\rm{L}_{\odot}$) and effective surface 
brightness ($10^{9.4}\,\rm{L_{\odot}\,kpc^{-2}}$)
are typical of LBGs \citep{heck05}.  An ACS far-UV continuum filter 
image of the center of \hoo\, is shown in Figure \ref{acs}.
Three separate starbursting nuclei are observed in the image. 
In deeper images, \hoo\,
appears to have an irregular morphology
suggestive of a merger.  This is consistent with  \ha~velocity
observations by \citet{ost01}
that show that \hoo\,is not dynamically relaxed.
IRAS measured a high dust temperature for
\hoo\, with 
$\rm{F_{60\micron}/F_{100\micron}}~\sim 1.3$ \citep{sand03}, implying an 
unusually intense UV radiation field associated with
a high star-formation rate per unit area. 
\citet{berg00} 
suggest that the lack of cold gas could imply that \hoo\,is
about to run out of the gas required to fuel the starburst. Lastly, it has
a very low gas phase metallicity of $\sim$20\% solar
\citep{berg02}. This low metallicity is presumably related
to the relatively low amount of dust extinction.  Galaxy properties of \hoo\, and, for comparison, \vv\, are
summarized in Table \ref{prop}.  While \vv\, has roughly four times the stellar mass and SFR, \hoo\, 
has a significantly higher UV surface brightness due to its smaller size.


\section{The Data}

\subsection{\fu~Observations\label{fusedata}}

\fu\, was launched on June 25th, 1999.  The \fu\,instrument is composed of 
four separate mirrors each having its own Rowland 
circle spectrograph \citep{moos00}.  Two mirrors
are coated with Al+LiF and two with SiC.  The LiF 
mirrors have better sensitivity in the $1000~\ang \lesssim \lambda  \lesssim 1180~\ang$
wavelength region, while the SiC are optimized for observations in the range
between $900~\ang$ to $1000~\ang$.  
During the period when our observations were taken,
pointing stability was $\sim0.5 \arcsec$.
The spectra are imaged onto
two micro-channel plate (MCP) detectors each having an A and B side.  
Each side produces 
both a LiF and SiC spectrum, for a total of 8 spectra
for every \fu\, observation.

Six time-tag mode exposures of \hoo\, were obtained through 
the LWRS aperture ($30\arcsec \times 30\arcsec$) on October 12, 2001.  
We have followed the standard data reduction methods for low S/N spectra.
The raw exposure data was run through the CalFUSE 3.1.8 data pipeline to produce
intermediate data (IDF) files and then the calibrated flux data.  As thermally-induced
rotations of the spectrograph gratings and mirror misalignment can cause
small zero point shifts in the wavelength calibration, we then cross correlate the
individual exposures.  For \hoo\, the largest zero-point wavelength
shift was very small at $ < 0.039$\,\AA.  These wavelength shifts were then retroactively applied to
the individual IDF files.  We  combined the IDF files for the individual exposures
to create one IDF file for each instrument channel.  
Extracting calibrated spectra from the longer, combined IDF files 
allows the CalFUSE pipeline to fit (and not just scale)
the various backgrounds.  
In order to minimize the detector background, we have extracted our
spectra using the smaller, point source detector regions
instead of the usual extended source regions \citep{dix07}.
This is possible as a visual examination of the
2-D detector image shows no evidence of galaxy emission outside
this smaller, point source extraction region.  
Although the exposure time is channel dependent,
the final exposure lengths are similar for all channels.
For reference,  the LiF 1A channel had 14953 seconds of 
total exposure time with 11940 seconds falling during orbital night.

\subsection{\ch~Observations\label{chandraobs}}

\hoo\, was observed by \ch's ACIS instrument on October 28, 2006.  
The target was observed in very faint mode with \hoo\, centered on the S3 aimpoint.
We  reprocessed the data with {\it ciao} 3.4.1 and followed the standard
\ch\, data reduction threads \citep{ciao}.  For spectra extraction we have used the {\it ciao}
command {\it specextract} and a local background from a region surrounding the galaxy.  
Approximately 1851 counts are observed from \hoo.
As described previously in \citet{grim05} we used a rescaled background
image included in {\it CALDB} 3.3.0.1 to create smoothed background subtracted
images.  The rescaled background image was also used to create an X-ray radial profile
of \hoo\, based on the 0.3 - 1.0 keV count data.

\section{Discussion}

\subsection{Physical Properties of the Hot X-ray Gas}

\begin{figure}
\centering
\leavevmode
\includegraphics[width=5in]{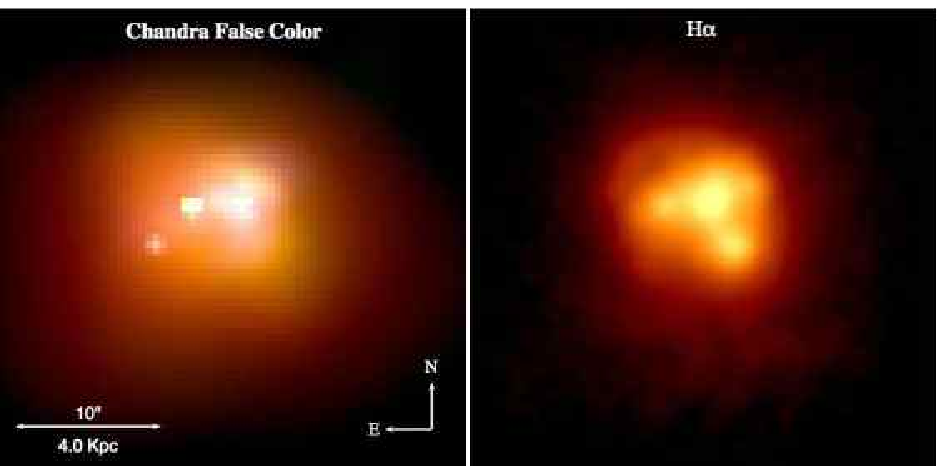}
\caption{
\ch\, false color (red 0.3-1.0 keV, green 1.0-2.0 keV, blue 2.0-8.0 keV) and \ha\, \citep{schmitt06} 
images of \hoo.
The two images have a similar morphology.  Knot B, likely the galaxy nucleus, is
the dominant feature in both the X-ray and \ha.  The \ch\, image suggests
a narrow band/disk of absorption running NE to SW through knot B.  This absorption feature
 also appears in the ACS FUV image (Figure \ref{acs}).
Knot D (SE of B), observed by \citet{kun03} in \lya\,emission, is only seen in the \ch\, image.  
\label{chandrafalse}}
\end{figure}

Figure  \ref{chandrafalse} is a three color adaptively smoothed X-ray image of \hoo\,
with 0.3-1.0 keV in red, 1.0-2.0 keV in green, and 2.0-8.0 keV in blue.  
Soft X-ray emission surrounds the galaxy and extends up to 6 kpc from
the galaxy.  The soft X-ray emission is also seen to have
a similar morphology to the \ha\,emission in the adjacent image \citep{schmitt06}.
  
Knot B shows the strongest emission in the X-ray and \ha\, images but is the weakest
knot in the UV.  \citet{kun03} observed that knot B is the reddest knot and exhibited
\lya\, absorption.  In both the X-ray and UV imaging data there appears to be an absorption
feature running across knot B from the NE to SW.  In all three wavebands Knot C appears
to be a single luminous star cluster while the UV image of knot A shows it
to be an assembly of several resolved star clusters.  This could explain knot C's 
relatively low X-ray surface brightness as compared to knots A and B.

A faint point source is visible SE of knot C in the X-ray image.
It is coincident with knot D which was identified by \cite{kun03} in their
\lya\,emission image of \hoo\, and therefore is unlikely to be a background
AGN.  We have extracted spectra of the point source using a local background region which
includes the diffuse thermal emission surrounding the source.
We detect  a total of $\sim 22$ counts from the source.  We fit the spectra with a powerlaw
and find $\Gamma\sim1.6$ and a luminosity of  
$\rm{L_{0.3-8.0\, keV}\sim2.2\times10^{39}\,erg\,s^{-1}}$.   The source's
luminosity and location near a star forming region suggest 
a possible ultraluminous X-ray source (ULX) identification but
are not conclusive.

\begin{figure}
\centering
\leavevmode
\includegraphics[width=3in,angle=-90]{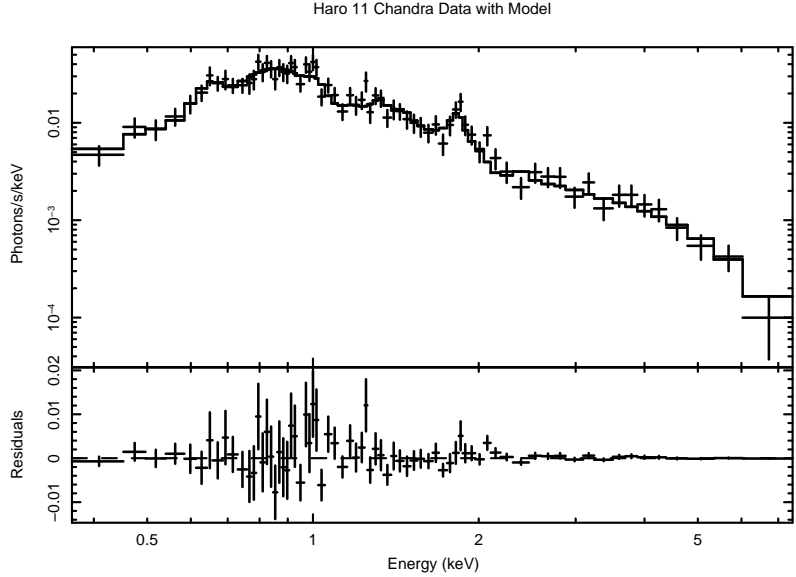}
\caption{
\ch\, spectrum and model of \hoo.  The model fit parameters can be found in 
Table \ref{xspecdata}.
\label{chandraspec}}
\end{figure}

The complete X-ray spectrum of \hoo\, is plotted in Figure \ref{chandraspec}.
As we have done in previous work \citep{grim05,grim06},
the spectrum was fit with an xspec model of \linebreak
$\rm{wabs(vmekal+zwabs\times powerlaw)}$.
 We find a hot gas temperature of
$0.68$ keV (see Table \ref{chandraspec} 
for the derived fit parameters).  Using
these fits we derive luminosities of \linebreak
$\rm{L_{0.3-2.0\,keV,thermal}\sim7.2\times10^{40}\,erg\,s^{-1}}$
for the soft thermal component and \linebreak
$\rm{L_{2.0-8.0\,keV,powerlaw}\sim1.0\times10^{41}\,erg\,s^{-1}}$
for the hard X-ray emission.  We estimate 
a SFR from the powerlaw fit to the hard X-ray emission of $25\,\rm{M_\odot\,yr^{-1}}$
using the relation of
\citet{franc03}.  This is in excellent agreement
with the \ha, FIR, and radio continuum derived values given  in \cite{schmitt06}.

\begin{figure}
\centering
\leavevmode
\includegraphics[width=2in,angle=-90]{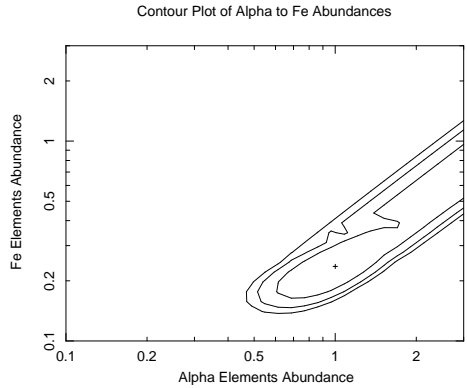}
\caption{
Contour plot (1-3 $\sigma$) of the variance in $\chi^2$ as a function of
\alphe\,and Fe abundances (relative to solar) from
the {\it xspec} fit to the X-ray spectrum.  Although it is impossible to 
determine the absolute abundances, 
the \alphe\, abundance (dominated by oxygen) in the hot gas
is significantly enhanced relative to the 20\% oxygen gas phase metallicity
derived in \citet{berg02}.     
\label{contour}}
\end{figure}

 We also find an enhanced \afe\, ratio 
 with a value of $\sim4$ times the solar ratio.
This is similar to the \afe\, ratios seen in in other starburst galaxies \citep[e.g. ][]{grim05}.
The enhanced \afe\, ratio is consistent with 
chemical enrichment of the hot gas by supernovae ejecta \citep{strick04b}.
Due to the usual degeneracy of the abundance parameters, 
it is impossible to place strong constraints
on the  absolute \alphe\, and Fe abundances.  
Figure \ref{contour} shows the 1-3 $\sigma$ contours of $\chi^2$ as a function of
\alphe\,and Fe abundances (relative to solar) from
our fit to the X-ray spectra.
The 20\% oxygen gas phase metallicity
derived in \citet{berg02} is significantly below
our allowed range for the  \alphe\, abundances, 
again suggesting SN enrichment of the hot gas.
Our best fit values for the absolute abundances of the hot gas are 
1 and .25 solar for the  \alphe s\, and Fe respectively.
If we assume an oxygen enrichment of $\sim 6$ (relative to solar) 
into the wind from supernovae \citep{lim06} we can estimate $\beta$, the ratio of total
mass added to the hot wind compared to that from stellar wind ejecta and SN 
\citep{strick05}.  Our results suggest a mass 
loading in the hot wind fluid of $\rm{\beta\sim7}$.

It is also worth estimating the mass of material entrained in the hot X-ray gas.
We can use the parameters for the fit to the total Chandra 
X-ray spectra of the diffuse gas in \hoo~
to derive estimates for the basic physical properties of the hot phase
of the wind. The normalization for the VMEKAL
component implies an emission integral (the volume integral of density
squared) of $5.6 \times 10^{63}$ cm$^{-3}$. For the geometrical volume
of the emitting region we take a sphere whose radius encompasses
90\% of the soft X-ray emission ($\sim$2.7 kpc). This then implies
a mean gas density of $n_e \sim 5\times 10^{-2} f^{-1/2}$ cm$^{-3}$  and
a gas mass of $M \sim 1 \times 10^{8} f^{1/2}$ M$_{\odot}$
(where $f$ is the volume filling factor of the X-ray emitting material).
For kT = 0.68 keV
the mean thermal pressure is $P = 1.1 \times 10^{-10} f^{-1/2}$ dyne cm$^{-2}$
and the total thermal energy content of the hot gas is $E = 2.5 \times 10^{56}
f^{1/2}\,\rm{erg}$.

Taking the characteristic timescale to be the above radius (2.7 kpc) divided by the
sound speed in the hot gas ($\sim$ 450 \kms) yields an outflow age of 6 Myr. The implied
outflow rates in the hot gas are then $\dot{M} \sim 18 f^{1/2}$ M$_{\odot}\,\rm{year}^{-1}$ and
$\dot{E} \sim 1.3 \times 10^{42} f^{1/2}$ \ergs. 
Since this neglects the kinetic energy in the wind material
it probably underestimates the total energy transport rate by a
factor of a few \citep{strick00}.
For volume filling factors similar
to those estimated for other starburst winds \citep[$f \sim$ 0.1 to 1, ][]{strick00b},
the implied mass outflow rate in \hoo~is thus comparable
to the total SFR of $\sim$ 25 M$_{\odot}\,\rm{year}^{-1}$ \citep{schmitt06}.
The rate of energy transport is similar to the total
rate at which we estimate that mechanical energy 
would be supplied by supernovae and stellar winds
in the starburst of $\sim 5 \times 10^{42} $ \ergs. 
This value was calculated using Starburst99 v5.1
assuming a Salpeter IMF from $1-100\,\rm{M_\odot}$ \citep{vaz05}.
These outflow rates significantly exceed
the rates for the cooler gas derived from the \fu\, data below, but 
are typical of winds in powerful local
starburst galaxies \citep[e.g.][]{heck03}. 

The Starburst99 calculations estimate a supernovae outflow
rate of $\rm{\dot{M}_{SN}\sim3.8\, M_{\odot}\,\rm{year}^{-1}}$.  
From the outflow rate of hot gas of 
$\dot{M} \sim 18 f^{1/2}$ M$_{\odot}\,\rm{year}^{-1}$,
we find a mass loading of $\beta \sim5\,f^{1/2}$.    This value agrees well
with our previous estimate of $\rm{\beta\sim7}$.
 It is also worth noting that the 
pressure and density derived for \hoo\, are almost identical to those in \vv\, 
with $P = 9.2 \times 10^{-11} f^{-1/2}$ dyne cm$^{-2}$ and
$n_e \sim 5\times 10^{-2} f^{-1/2}$ cm$^{-3}$  \citep{grim06}.

\subsection{\fu Observations of the Wind}
\begin{figure}
\centering
\leavevmode
\includegraphics[width=4in,angle=90]{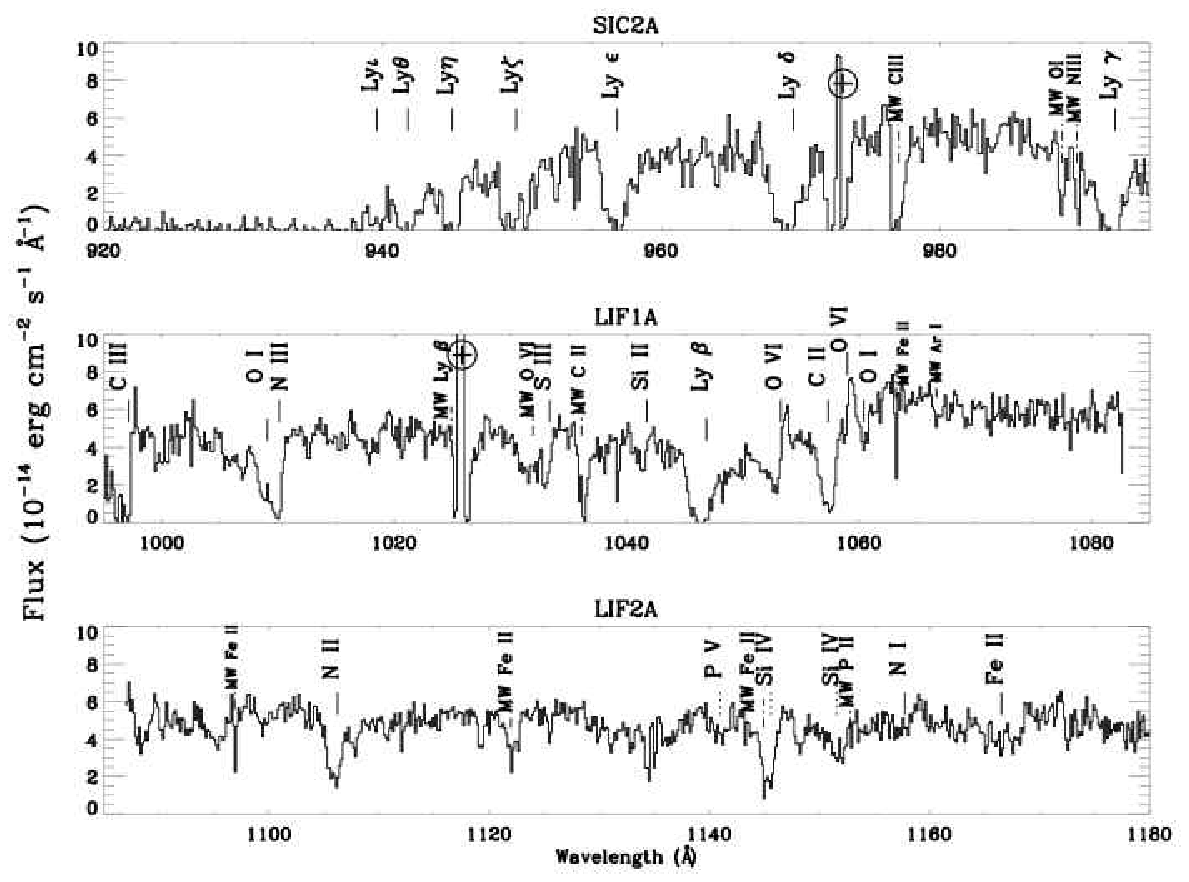}
\caption{
SiC 2A, LiF 1A, and LiF 2A spectra of \hoo\, covering the majority
of the \fu\, spectral range.  Prominent \hoo\,absorption lines, milky way
lines, and airglow features (earth symbol) have been identified.
\label{fullspectra}}
\end{figure}

The \fu\, data reveals a large number of absorption features, as shown in Figure  
\ref{fullspectra}.
These absorption lines probe the stellar population, inter-stellar medium (ISM), 
and outflowing gas of \hoo.  We have fit absorption lines to most of the most 
promiment features using the IRAF 
\citep{tod96}
 tool {\it specfit} \citep{kriss94}.  Each line was fit using a freely varying powerlaw for 
 the continuum and a symmetric gaussian absorption line.  When absorption lines 
 are  blended (e.g. \ion{O}{1}~$\lambda$989 and \ion{N}{3}~$\lambda$990), we have 
 fit them simultaneously.  For the majority of absorption lines we have used the LiF 
 channels to make two independent measurements of the equivalent width, velocity, and 
full-width half-maximum (FWHM).  For lines with only one
measurement, the line either fell in the gap between LiF 
channels or was in the region (1125-1160\AA) 
affected by partial light blocking by a wire grid (e.g. the worm) on the LiF 1B channel.  We have ignored the SiC channels above 1000 \AA\,
due to their lower S/N.  
 Tables \ref{absdata} and \ref{phabsdata} display
the results of our fits to the data for the ISM lines and stellar photospheric lines respectively.  
Errors are $1\sigma$ and are
calculated from the minimization error matrix which is 
re-scaled by the reduced $\chi^2$ value 
\citep{kriss94}.

We use the stellar photospheric lines as an independent measurement of the
system velocity of \hoo.  Previous work by \citet{berg06} used \ion{Ca}{2} absorption lines
to derive a velocity of 6180 \kms.  This compares favorably with our own measurements of 
6163 - 6202 \kms\, from \ion{Si}{4} and \ion{P}{5} stellar lines.  Therefore we will
use the value of 6180 \kms\, as the galaxy's systemic velocity.

\begin{figure}
\centering
\leavevmode
\includegraphics[width=5in,angle=90]{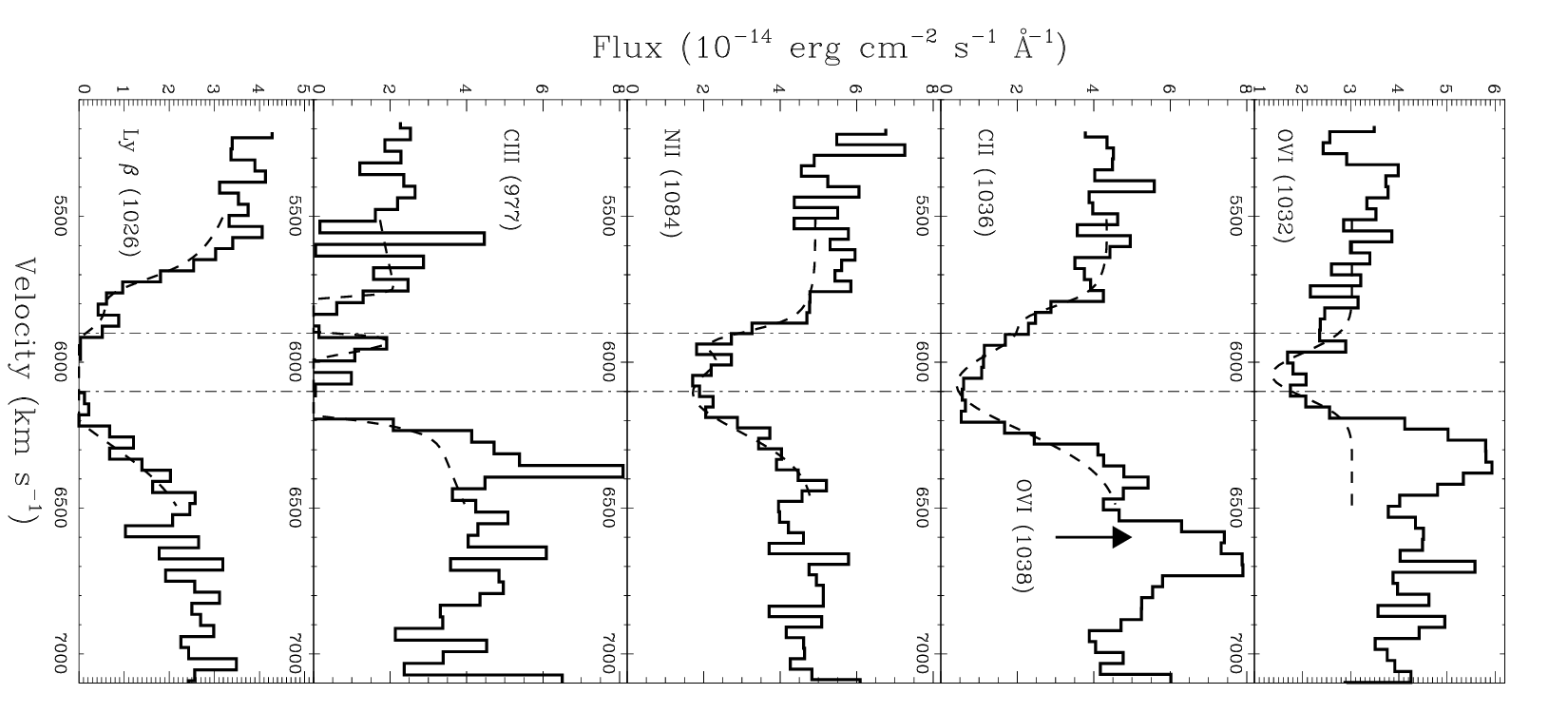}
\caption{
Expanded plots of five strong absorption features in the \fu\,spectra
of \hoo.  All of the absorption lines except 
\ion{O}{6} show signs of one strong, broad absorption line at $\sim6100$ \kms\,
and another weaker absorption line at $\sim5900$ \kms.  
Both are significantly blueshifted relative to
the galaxy systemic velocity of $6180$ \kms.
We have overlaid two component absorption line fits for the
\ion{N}{2}, \ion{C}{3}, \ion{C}{2}, and $\rm{Ly\,\beta}$ lines and
vertical lines at $5900$ and $6100$ \kms\, for visual comparison.  
The absorption feature at $5900$ \kms\, is clearly visible in the spectra
although there is some variability in the centroid velocity.
The \ion{O}{6}~$\lambda$1032
line has a P-Cygni profile as it has absorption on the blue wing and emission
on the red side.  For
\ion{O}{6} we have plotted a single absorption line although it does not
well represent its non gaussian profile.  \ion{O}{6}~$\lambda$1038
emission (and possibly absorption) is  observed in the \ion{C}{2}~$\lambda$1036 plot.  
\label{absprofiles}}
\end{figure}

 The ISM lines in Table \ref{absdata} suggest a centroid blueshift relative to the
 galaxy of $\sim100$ \kms.  A closer examination of
 the line profiles shows that they are consistently asymmetric.  
 This mild asymmetry in the line profile can be seen in Figure \ref{absprofiles}.  
 There appears to be excess absorption on the blue side.  The gaussian models are 
 overestimating the FWHM in order to fit the absorption line profiles.  In keeping 
 with our previous work on \vv,
 we have fit the strongest absorption lines with a second, blue absorption line.  The results of our 
 two-component fits are listed in Table \ref{twoabsdata}.
 
We associate the stronger absorption feature in our fits with the host galaxy or a low
velocity outflow.  In general, the
centroids of these lines are at $\sim6100$\,\kms.  The central velocity is 
blueshifted $\sim80$\,\kms\, relative to the host galaxy.  The stronger absorption
lines have FWHMs of $\sim300$\,\kms, suggesting that the  lines do
include the galaxy's ISM.
The weaker absorption lines are generally at $\sim5900$\,\kms. 
We identify the second set of absorption lines with a high velocity outflow of material
 moving at $200-280$\,\kms\, away from the galaxy.  Interestingly, a 
 comparison of the equivalent widths of \ion{C}{2} to \ion{C}{3} and \ion{N}{2} to  \ion{N}{3} show that 
 the higher ionization lines are relatively stronger in the outflow as compared to the 
 lower velocity absorption features.  This could be a product
 of shocks between the outflow and the ambient material.


In Figure \ref{absprofiles} it is clear that the 
coronal gas line \ovi$\lambda1032$ has a different profile 
than the other ISM absorption lines.  
While the \ovi\,centroid velocity of $6043\pm34$\,\kms\,
is roughly consistent with those of the ISM absorption lines,
there are significant discrepancies.
First, a single absorption
line fits  \ovi\, poorly,  due to a weak blue wing
extending several hundred \kms\, away from the
 galaxy.  While this blue wing is relatively weak
 compared to the primary feature, it extends to much lower
 velocities (e.g. higher outflow velocities) than the 
 blue wings we observe in some of the
 other absorption profiles.
Secondly and more intriguingly, \ovi\,
 emission appears to be present, forming part of a 
 P-Cygni profile.  An emission feature is observed 
 at $6303\pm21$\,\kms.  This identification is supported by a 
 corresponding emission feature for \ovi$\lambda1038$
 with a similar centroid at $6282\pm30$\,\kms.  
They also have similar FWHMs of
$185\pm47$ and $151\pm55$ \kms\,for \ovi$\lambda1032$
and \ovi$\lambda1038$ respectively.  
 An absorption feature
 is not observed for \ovi$\lambda1038$ but it is likely hidden in the
 broad \ion{C}{2}$\lambda1036$ absorption feature.  
 In fact, the previously described two component absorption line profile
 for \ion{C}{2} fits the red wing particularly poorly (Figure \ref{absprofiles}) and has a large
FWHM, consistent with  contamination by an \ovi$\lambda1038$ 
 absorption feature.  The \ovi$\lambda\lambda1032,1038$ doublet
 is the only ion that we unambiguously identify in emission
  although there is a possible emission feature for \ion{C}{3}$\lambda977$
 at 6356\,\kms.  The \ovi\, P-Cygni profile has been previously identified by
 \citet{berg06}.
 
 
Broad P-Cygni profiles
are common for \ovi\, in spectra of O stars \citep{pell02}.
While a P-Cygni profile has  not been observed in other starburst galaxies,
the high UV surface brightness, low UV attenuation,
and high SFR of \hoo\,could suggest that we are detecting an aggregate
stellar profile.  
However, the observed \ovi\,emission and absorption features are significantly
narrower than those seen in stellar spectra.  
In particular, narrow \ovi$\lambda\lambda1032,1038$ emission is not observed
in the stars in the  sample of  \citet{pell02}.  The narrow \ovi\,feature is
also inconsistent with the synthetic starburst spectra generated by \citet{rob03}.
We therefore explore the possibility that the \ovi\, profile is not a stellar feature
but is instead due to interstellar gas.

In previous analyses of starburst galaxies 
\citep[e.g.][]{heck01,grim06} the \ovi\,absorption has been the most
blueshifted and broadest feature.  
The extreme outflow velocities of the \ovi\,absorption suggests
that in those galaxies, \ovi\, is not produced 
directly by the cooler outflowing clouds responsible
for the other absorption lines.   Instead, \citet{heck01} attribute the 
production of  \ovi~
to the intermediate temperature regions created by the 
hydrodynamical interaction between hot outrushing gas
and the cool fragments of the ruptured superbubble seen 
in H$\alpha$ images. Such a situation is predicted to be
created as an overpressured superbubble 
accelerates and then fragments as it expands out of the galaxy \citep{heck01}.  

\citet{heck02} derive a simple and general relationship between the \ovi~
column density and absorption line width which holds whenever 
there is a radiatively cooling gas flow passing through the coronal
temperature regime.  They showed that this simple model 
accounted for the properties of \ovi ~ absorption line systems as 
diverse as clouds in the disk and halo of the
Milky Way, high velocity clouds, the Magellanic Clouds, 
starburst outflows, and some of the clouds in the IGM (but see Tripp et al. 2007,
in prep, for a more thorough test of the relation for IGM clouds).  
From our fit to the \ovi$\lambda1032$ absorption feature we have derived
a column density of $\rm{N_{OVI}=4.2\times10^{14}}\,cm^{-2}$.
\hoo\, does fall within the scatter in the relationship, but we would expect
a slightly higher \ovi\,column density
for the measured line width of $\rm{\sqrt{2}\sigma=b=106\,\ks}$.
While the equivalent width and breadth of the \ovi\,absorption  are 
roughly consistent with \ovi\, production in
the interface between the outrushing gas and the 
cool shell fragments \citep{heck01}, the kinematics of \ovi\,
are inconclusive.  Specifically, unlike our previous 
observations of \vv\, and NGC\,1705 \citep{grim06,heck01}, 
the primary \ovi\, absorption feature is not blueshifted relative to the cooler, ISM 
lines.

 

 \ovi\, emission has previously been observed in only
 two spiral galaxies, NGC\,4631 \citep{ott03} and our own
 Milky Way \citep[][and references therein]{dix06}.  The \ovi\, emission in the
 halo of NGC\,4631 was attributed to cooling gas from
 a galactic chimney \citep{ott03}.  Similarly,
 for \hoo\,, we could attribute the \ovi\, emission
to radiatively cooling gas
 associated with the outflow. 
 As \ovi\, is the dominant coolant for gas at temperatures 
 $\rm{T\sim10^{5.5}\,K}$, 
 \ovi\,cooling could inhibit the ability of the outflowing gas to
escape the galaxy.  Therefore it is interesting 
to estimate the implied energy losses associated with
the \ovi\, emission.

The observed flux in \ovi\, emission is
$2.3\pm0.8\times10^{-14}\,\rm{erg\,s^{-1}\,cm^{-2}}$ and
$1.4\pm0.5\times10^{-14}\,\rm{erg\,s^{-1}\,cm^{-2}}$
for the $\lambda1032$ and $\lambda1038$ lines respectively.
While the \ovi$\lambda1038$ flux could be an underestimate due to
\ion{O}{1}$\lambda1039$ absorption, these measured values
are near the theoretical upper limit
of $\rm{F_{OVI\lambda1032}/F_{OVI\lambda1038}}=2$
found in an optically thin gas.
We correct the fluxes for both 
galactic extinction and intrinisic
dust attenuation.
To correct for galactic extinction we 
use E(B-V)=0.011 \citep{schleg98} and assume
the extinction law of \citet{card89}. 
For intrinsic attenuation we start with 
work by \citet{seib05}.  They used
{\it GALEX} and {\it IRAS} data to
study the relationship between
UV attenuation (1500 \AA) and the IR excess.
From Table \ref{prop}
we find an IR excess for \hoo\, of $\rm{\log(L_{FIR}/L_{UV}})= 0.53$.  This IR
excess correspond to $\rm{A_{1500\,\AA}\sim1.6\,mag}$ \citep{seib05}.
We then extend this dust attenuation to 1035 \AA\,
following the work by \citet{leith02}. 
They used Hopkins Ultraviolet Telescope (HUT) 
observations of star-forming galaxies to study 
dust attenuation at shorter wavelengths.
They found a factor of 1.3 increase in
the dust attenuation from 1500 to 1035 \AA.
We therefore calculate $\rm{A_{1035\,\AA}\sim2.1\, mag}$. 
Finally we estimate a total corrected \ovi~luminosity
of $\sim2.8\times10^{41}$ \ergs.
\ovi\, is responsible for $\sim30\%$ of the
cooling in coronal gas \citep{heck01} so 
$\rm{L_{coronal}\sim9.3\times10^{41}}$ \ergs.
This luminosity is larger
than \hoo's thermal X-ray luminosity ($7.2 \times 10^{40}$ \ergs) 
and
is comparable to both the energy outflow rate of
the hot gas ($\dot{E} \sim 1.3 \times 10^{42} f^{1/2}$ \ergs)
and the
mechanical energy being input by
supernovae ($\sim 5 \times 10^{42} $ \ergs).  
Therefore, if the \ovi\, emission is 
produced by radiative cooling of
the outflowing gas, it suggests that a significant
percentage (20\%) of the energy input from the supernovae could
be radiated away by the coronal gas.


Several previous works have discussed the  $\rm{L_{OVI}/L_{X-ray}}$ ratio.
In their detection of \ovi\, emission in NGC\,4631, \citet{ott03} estimated 
$\rm{L_{OVI}/L_{X-ray}}\sim0.11$.  
Although \ovi\, emission was not observed in NGC\,1705 and M82, lower limits
of   $\rm{L_{OVI}/L_{X-ray}}\lesssim3$ \citep{heck02} and $\lesssim0.4$ \citep{hoop03} 
were determined respectively.  Our own result, $\rm{L_{OVI}/L_{X-ray}}\sim3.9$ seems
anomalously high in comparison to these previous results.

It is possible that the \ovi\, emission line in \hoo\, is not tracing the
rate at which coronal gas is radiatively cooling. Instead it may arise
through resonant scattering of far-UV continuum photons from the starburst
off \ovi\, ions in the outflow. The redshift of the emission component is a
natural consequence of such a picture (this is a classic P-Cygni profile).
There is a shell of coronal gas flowing out from the starburst, and is
traced by \ovi\, ions. The front side of the flow is seen as blue-shifted
absorption against the far-UV continuum of the starburst, while resonant
scattering by \ovi\, ions on the back side of the flow produces a red-shifted
emission line. For a symmetric flow geometry, and in the absence of
significant differential dust extinction of the OVI emission from the back
side, this would predict that the equivalent widths of the blueshifted
absorption and redshifted emission lines would be equal and opposite. This
is roughly what is observed in Haro 11.

If this is the correct interpretation, the question is then why a redshifted
emission component is not seen in the other ions (lower ionization states).
This would have to mean that the \ovi\, and cooler gas are not co-spatial:
redshifted emission from the back side of an outflow is obscured by dust for
the low ions, but not for \ovi\,. This could be the case (for instance) if the
size scale of the flow were significantly larger in OVI so that more of the
back side lay beyond the region of significant dust extinction. 

With the present data it is not possible to determine whether the \ovi\,
emission is produced by radiative cooling, or resonant scattering. If it's
the latter, then the true contribution to radiative cooling by the coronal
gas discussed above must be regarded as an upper limit. This reinforces the
idea that radiative cooling is not significantly influencing the dynamics of
the wind.

\subsection{Escape of Ionizing Radiation?}

Recent observations imply that the epoch of reionization 
occurred at a redshifts $\rm{6<z<20}$ 
(Becker et al. 2001, Fan et al. 2002, Spergel et al. 2003). Several mechanisms have been proposed
for reionization including UV photons from star formation or active galactic nuclei (AGN), 
X-ray photons 
from supernovae and mini-quasars (Oh 2001), and even sterile neutrino decay 
(Hansen \& Haiman 2004).  However X-ray background and cosmic microwave 
background (CMB) observations
rule out significant contributions by X-ray photons and neutrino decays 
(Mapelli \& Ferrara 2005; Dijkstra et al. 2004). UV photons from quasars
likewise appear to be inadequate due to the rapidly
falling space density of quasars at $\rm{z>3}$ (Fan et al. 2001).
While recent work suggests that
UV photons from stars could reionize the universe 
(Yan \& Windhorst 2004), this inference is quite uncertain. First, the cosmic star formation
at the relevant redshifts is still uncertain. Second, the fraction of ionizing photons that
are able to escape to the IGM ($f_{esc}$) is even more uncertain.

The escape of ionizing continuum radiation from star forming galaxies 
was first observed by Steidel et al. (2001) in a sample of 
$\rm{z\sim 3.4}$ LBGs.  
It has been suggested that the lower
luminosity LBGs (e.g. $\rm{L < L_*}$) could play an
important role in reionization (Bouwens et al. 2006).
Galactic winds appear to be a global property of LBGs
(Shapley et al. 2003), and this may explain why a significant fraction of their ionizing radiation
is able to escape.

More recent work by \citet{shap06} focused on deep rest-frame UV spectroscopy of
a sample of 14 LBGs.  They detected significant 
Lyman continuum emission in only two of the galaxies.  
They found a sample-averaged relative Lyman continuum
escape fraction of 14\% but were unable to find
a galaxy property that controlled
the detection of the
Lyman continuum. Their results suggest that
even at high redshifts ($\rm{z\sim3}$) significant Lyman leakage is
 the exception.

Most recently \citet{berg06} published an analysis showing 
the detection of Lyman continuum emission in \hoo.  \hoo's low 
neutral hydrogen mass limit of $\lesssim10^8\rm{M}_\odot$ \citep{berg06}, 
high FUV surface brightness, irregular morphology, and low metallicity \citep{berg02} all suggest
that \hoo\, is an ideal local candidate for  Lyman continuum emission.  
\hoo's Lyman continuum leakage was in fact our prime motivation
for obtaining \ch\, data of the galaxy.  As part of our analysis we have
re-analyzed the \fu\, data and have come to a different conclusion than the
previous work by \citep{berg06}.

\begin{figure}
\centering
\leavevmode
\includegraphics[width=3in,angle=90]{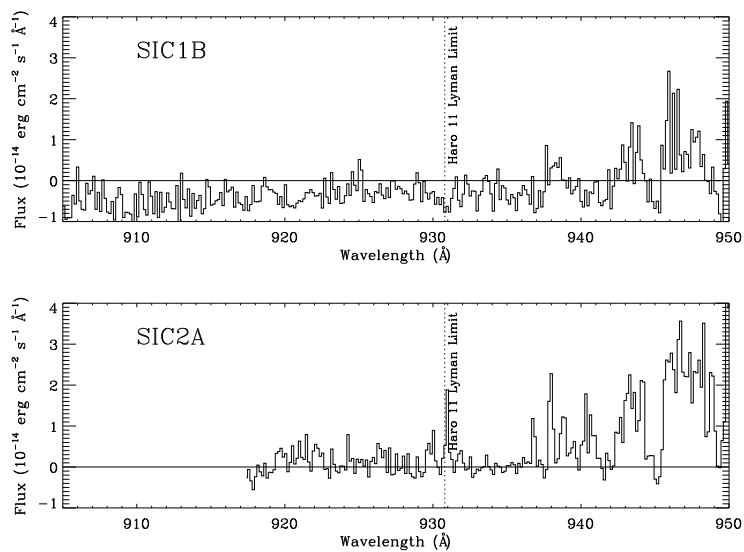}
\caption{
SiC 1B and SiC 2A night spectra (binned by 12 pixels=0.156\,\AA) of the Lyman continuum region.  Both spectra have
been run through the standard \fu\, data pipeline.  The SiC 1B background
has clearly been oversubtracted in this wavelength region resulting in a negative flux.
\label{sic1b2aorig}}
\end{figure}

\begin{figure}
\centering
\leavevmode
\includegraphics[width=4in]{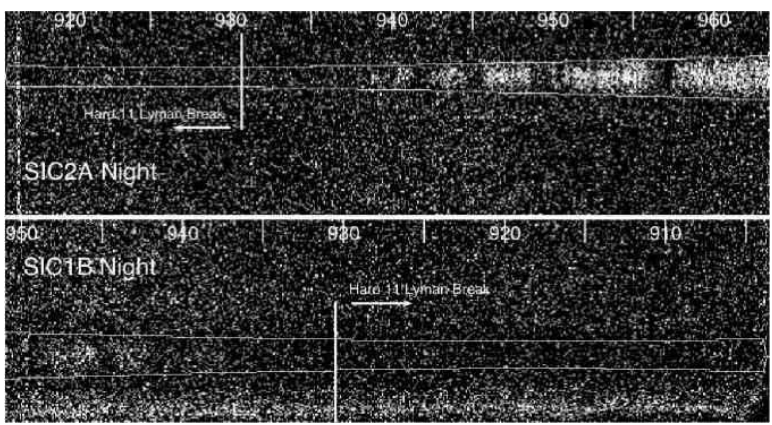}
\caption{
SiC 1B and SiC 2A image of the detectors in the Lyman continuum region during 
orbital night.  The 
point source extraction aperture has been overlayed for both detectors.  No clear
emission can be seen in the Lyman continuum region on either detector.  
Background subtraction of the SiC 1B spectra is complicated by its location near the lower  edge of 
the detector.  Vertical structure in the background can also been seen on both detectors.
\label{detector}}
\end{figure}

For the following discussion we  focus
primarily on the night only data.  While we lose 4 ks (12 ks total night exposure) of observing time,
the night background level is considerably lower and relatively well behaved. 
In Figure \ref{sic1b2aorig} we have plotted the extracted spectra for the SiC 2A and SiC 1B detectors.
As discussed in section \ref{fusedata} we have used the standard 
extraction methods  for 
low S/N spectra to produce these plots.   
The SiC 1B spectra show a negative flux, a common feature 
of low S/N LWRS  spectra at these wavelengths.  
This is related to the difficulty in fitting the background models to the LWRS aperture region
as it is near the bottom edge of the detector (see Figure \ref{detector}).  Unlike the SiC 1B spectra, the SiC 2A spectra have
 a small positive net flux.  

\begin{figure}
\centering
\leavevmode
\includegraphics[width=4in,angle=90]{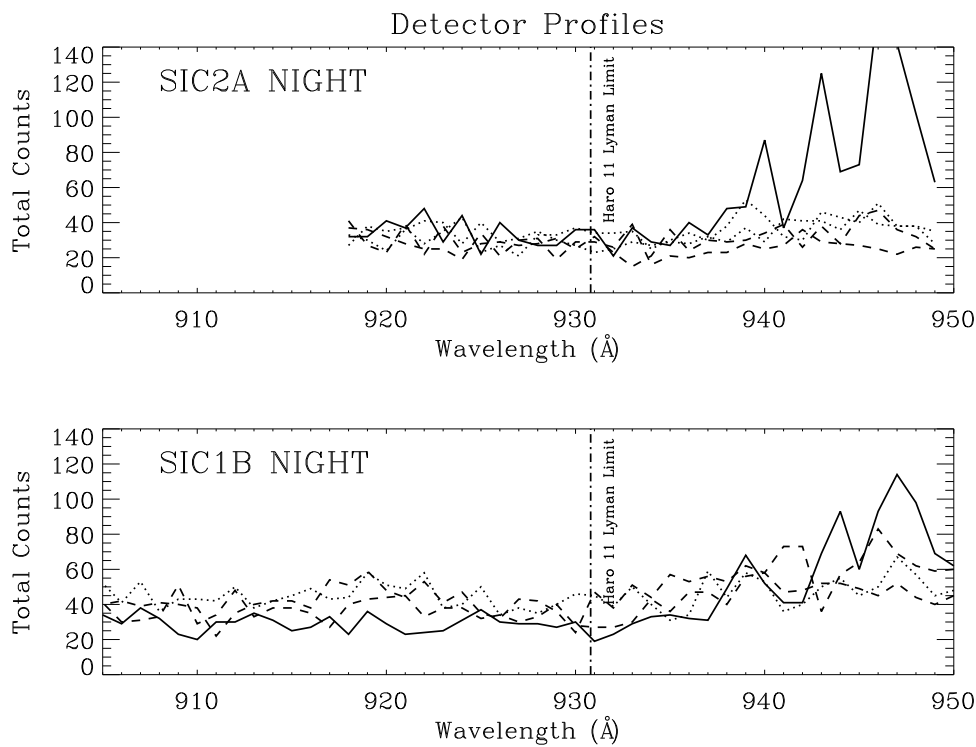}
\caption{
We have extracted the number of counts in 1\AA\, bins from a rectangular region
enclosing the source aperture (black line) and background regions directly
below (dotted) and directly above (dashed) the aperture.
The background profiles are very similar to the source profile 
below the Lyman limit.  At variance with \citet{berg06} there does not appear to be any excess counts
in the source regions below the Lyman limit.  Note that several Lyman series absorption
lines are visible above the Lyman limit.  These profiles are consistent with a visual
analysis of the  2-D spectra shown in Figure \ref{detector}. 
\label{radialprofile}}
\end{figure}

The oversubtracted  SiC 1B spectra suggests the relative importance of the background 
subtraction.  Figure \ref{detector} shows images of the two micro-channel plates (MCPs) in the regions
of interest.  We have overlaid the point source extraction region and wavelength scales on the image. 
While galactic emission is clearly visible at wavelengths longward of \hoo's Lyman break ($\sim 930.8\,\rm{\AA}$),
there is no convincing visual evidence of Lyman continuum emission in these images.  More importantly, possible
emission features in SiC 2A are not replicated in the SiC 1B image.  
In order to examine the spectra
more directly we have extracted
the total counts in 1\AA\, bins for a rectangular region approximating the source aperture.  We have also
extracted the same information from adjacent regions of the same size directly above and below the source aperture (Figure \ref{radialprofile}).  These background  profiles are 
similar to the source spectrum 
for wavelengths below the
Lyman limit.  If present, any Lyman continuum flux 
appears to be fainter than the variance present in the background regions.

\begin{figure}
\centering
\leavevmode
\includegraphics[width=4in]{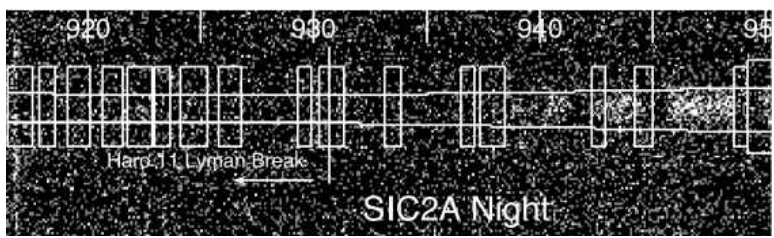}
\caption{
SiC 2A night image of the detector with the SiC 2A airglow regions identified by CalFUSE 3.1.8.  
\label{sic2aairglow}}
\end{figure}

In Figure \ref{sic2aairglow} we show
the regions flagged by the CalFUSE pipeline as 
possibly contaminated by geocoronal or scattered solar light.  
While these features are significantly weaker during orbital night, the region of interest between 
$920 - 930 \rm{\AA}$ is almost fully covered by these features.  
Determining the background in these regions is thus particularly difficult.
Regions of background contamination were interpolated across when 
the \fu\, scattered light background files
were created \citep{dix07}.

\begin{figure}
\centering
\leavevmode
\includegraphics[width=3in,angle=90]{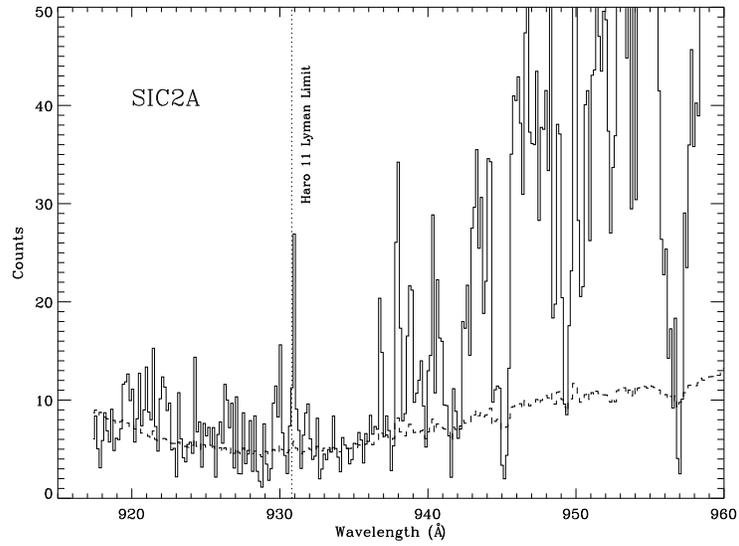}
\caption{
Total counts (spectra + background) from optimal extraction 
(binned by 12 pixels=0.156\,\AA) in the SiC 2A night time observation as a function of wavelength.  The
fitted background is shown as a dashed line.  This plot again shows that the background
dominates any signal in the Lyman continuum region.
\label{sic2acounts}}
\end{figure}

There are several components to the \fu\, background.  The intrinsic background of
$\sim\, 0.5\, \rm{counts\, cm^{-2}\, s^{-1}}$ consists of beta decays from $^{40}\rm{K}$ on the MCP glass  
and cosmic rays \citep{dix07}.  In addition, a contribution from geocoronal light varies from
20-300\% of the intrinsic background during the orbit.  The \fu\, pipeline uses independent
measurements of the dark count, night geocoronal scattered light, and day geocoronal scattered light
to fit the background observed on {\bf unilluminated} regions of the detector.  An 
optimal extraction algorithm then extracts the source counts
from the background \citep{horn86}.  Figure \ref{sic2acounts} shows
the total counts and fitted background model for the SiC 2A night time data.
The background clearly dominates the observed counts in the regions below
the Lyman limit.  While the fitted background is generally reasonable, it does
over-subtract in some regions.  This again highlights the difficulties in subtracting
a temporally and spatially varying background from relatively low S/N data.

We have derived total fitted source and background fluxes for the SiC 2A night data
in two wavelength ranges below the Lyman limit.  From 
920-925 \AA\, we found a
source flux of $1.9\times10^{-15}\,\rm{erg\,cm^{-2}\,s^{-1}\,\AA^{-1}}$ 
compared to a background
flux of $8.7\times10^{-15}\,\rm{erg\,cm^{-2}\,s^{-1}\,\AA^{-1}}$.  
For comparison, we also examined
the 925-930 \AA\, region within which we would expect less airglow contamination (e.g. Figure \ref{sic2aairglow}).
This region has a significantly lower source flux of 
$5.6\times10^{-16}\,\rm{erg\,cm^{-2}\,s^{-1}\,\AA^{-1}}$
with a comparable background of $6.1\times10^{-15}\,\rm{erg\,cm^{-2}\,s^{-1}\,\AA^{-1}}$.  
The source flux varies by a factor of $\sim 3$ between these two regions although we would not expect
the Lyman continuum to vary strongly \citep{berg06}.  
A 20\% error in the background subtraction would erase any source flux even in the 
920-925 \AA\, region.  

\begin{figure}
\centering
\leavevmode
\includegraphics[width=3in,angle=90]{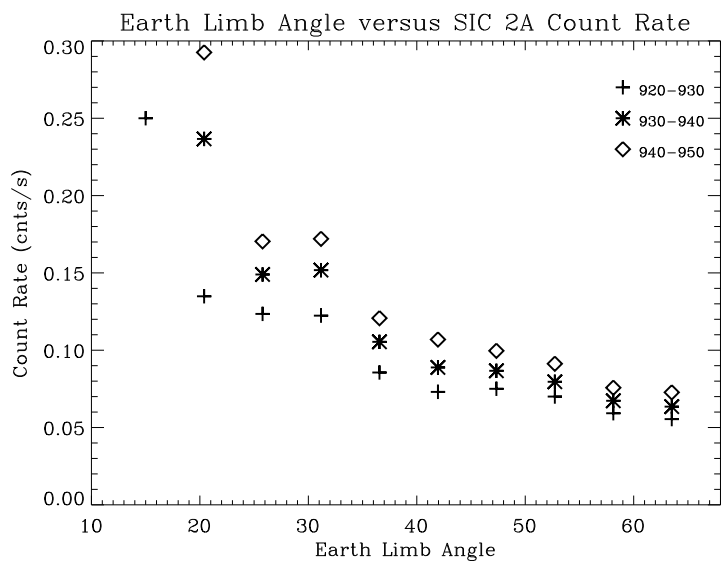}
\caption{
This plot shows the count rate in three wavelength regions of the LWRS aperture for the 
SiC 2A detector as a function of earth limb angle.  
The 920-930\,\AA~ bandpass, which represents the Lyman continuum emission region, 
has the lowest count rate for all limb angles and appears to still be 
falling at the highest angles.  The expected intrinsic count rate of the detector from 
cosmic rays and $^{40}\rm{K}$ decays is $\sim 0.05$ counts $\rm{s^{-1}}$ which
 is just below the count rate seen at the highest
earth limb angles.  This suggests that any excess emission in the SiC 2A 
spectra below the Lyman limit can be attributed to airglow.
\label{earthlimb}}
\end{figure}

As we have stated, the \fu\, background varies significantly
during an observation.  Airglow and geocoronal scattered light
both contribute to the background and vary depending on
the pointing of the satellite relative to the earth.  
As the \fu\,data is time-resolved, we can study
the detector count rate as a function of wavelength and
earth limb angle.  In Figure \ref{earthlimb} we have
plotted the earth limb angle versus detector count rate 
for three wavelength regions.
This plot includes both night and day data
so that we could include the lowest earth limb angles.  
The Lyman continuum region is represented by
the 920-930 \AA\, points while the 930-940 \AA\, and
940-940 \AA\, regions contain some emission intrinsic to
\hoo.  The Lyman continuum region
has the lowest count rate for all earth limb angles.
More importantly, as the earth limb angle increases, the count rate 
continues to fall even at the highest
earth limb angles.  We estimate an expected dark count rate from cosmic rays and $^{40}\rm{K}$ 
of $\sim0.05\,\rm{counts\,s^{-1}}$  which closely matches the count rate in the Lyman
continuum region observed at the 
highest earth limb angles.  This again suggests that the observed Lyman continuum emission
is likely a systematic artifact of extracting low S/N spectra from a dominant
and variable background.

\begin{figure}
\centering
\leavevmode
\includegraphics[width=3in,angle=90]{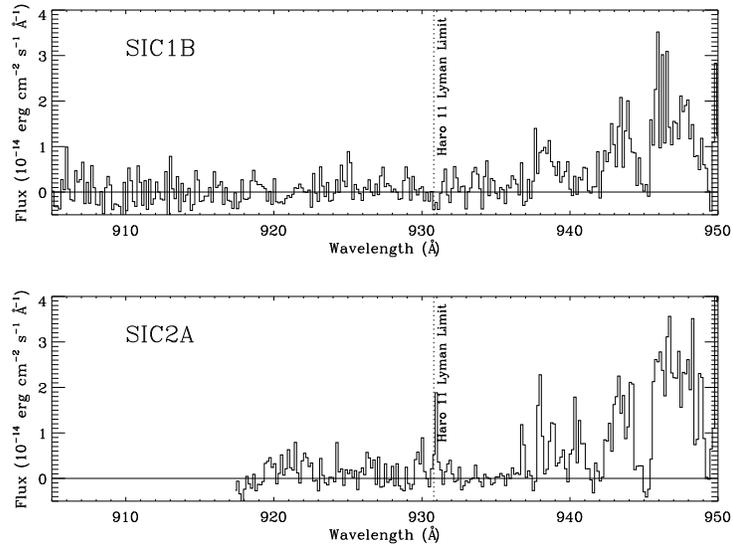}
\caption{
SiC 1B and SiC 2A night spectra (binned by 12 pixels=0.156\,\AA) of the Lyman continuum region.  The background subtraction
of the SiC 1B spectrum has been adjusted to match a flux level similar to that in the
SiC 2A spectrum.  Neither spectra show convincing evidence of Lyman continuum
emission.  In particular it should be noted that apparent features in one spectra
are not replicated in the other as expected in noisy, low S/N spectra.
\label{sic1b2afix}}
\end{figure}

For Figure \ref{sic1b2afix} we have rescaled the SiC 1B background subtraction to
match the observed SiC 2A spectra.    In comparing the two spectra it is worthwhile
to note that in the regions below 930\,\AA\,, possible spectral features are not replicated
in both spectra.  This also suggests that the detected emission is largely a combination
of detector noise and systematic effects from the background subtraction.

A comparison to the results by \cite{berg06} of the SiC 2A data
show a positive offset of the continuum level by  
$\sim1\times10^{-14}\,\rm{erg\,cm^{-2}\,s^{-1}\,\AA^{-1}}$.  
The continuum shapes however are similar.  
In the \citet{berg06} spectra it is worthwhile noting that none of the
absorption lines are black.  This includes two narrow Milky
Way absorption lines \ion{C}{3} $\lambda 977$ and \ion{C}{2} $\lambda 1036$
which would be expected to be saturated at line center, independent of
the properties of \hoo.  These lines are black in our spectra as seen in
Figure \ref{fullspectra}.

As we are analyzing the same observational data, we have three 
possible reasons  to explain the discrepancy between our results and those of \citet{berg06}.  
First, we are using a 
newer version of the CalFUSE pipeline (3.1.8 versus 3.0) which includes
several improvements in the background analysis.  In particular
v3.0 did not properly account for regions of the detector excluded because of 
airglow contamination and therefore frequently underestimated the total background.
Secondly, we suspect that the extended source extraction region was used for the
prior analysis by \citet{berg06}.  For \hoo\, the point source extraction region includes all of the 
source flux and has
a higher S/N due to the smaller extraction region on the detector.  Lastly, it is possible that the
individual exposures were separately background subtracted.  For low S/N spectra
it is preferable to combine the exposures and allow the CalFUSE pipeline to fit
a background model to the merged dataset.  If the background subtraction
is performed on the individual exposures of \hoo\,, the CalFUSE pipeline will
simply scale the subtracted background by the exposure time.  As the background
is varying both spatially and temporally, this causes 
systematic offsets in the background subtraction.  This explanation is likely
as it would also explain the offset seen even in the relatively high S/N LiF 1A spectra 
of \citet{berg06}.

Although we find no convincing evidence for Lyman continuum emission it is useful
to derive an upper limit.  Following \citet{berg06} we use the  920-925 \AA\, flux 
and correct for galactic extinction.  We derive
$\rm{f_{900}\lesssim2.3\pm0.7\times\,10^{-15}\, erg\,cm^{-2}\,s^{-1}\,\AA^{-1}}$
as compared to the previous result of 
$\rm{f_{900}=1.1\pm0.1\times\,10^{-14}\, erg\,cm^{-2}\,s^{-1}\,\AA^{-1}}$
\citep{berg06}.  Our $1\sigma$ errors are purely statistical and do not include the
dominant systematic errors.  Using our derived value of $\rm{f_{900}}$
and the formulism for deriving $\rm{f_{esc}}$ derived by
\citet{berg06}, we find $\rm{f_{esc}\lesssim2\%}$.


\subsection{X-ray Properties of LBG Analogues}

In \citet{grim05}  we analyzed \ch\, observations of the
diffuse emission in 23 star-forming galaxies.  We showed that the properties 
of the hot gas  (e.g.
temperature, metal enrichment, and luminosity)  were remarkably 
consistent throughout the sample.  These findings suggested that the
same physical mechanism was
producing the diffuse X-ray emission throughout the sample.
We found that shocks, driven by a galactic ``superwind'' powered by the kinetic
energy collectively supplied by stellar winds and supernovae in
the starburst,  were the likely explanation for this correspondence. 
As we have now obtained deep \ch\, observations of two
LBG analogues, it is worth examining how they fit within
this framework.
While it is 'dangerous' to make
general conclusions based on such a small sample size,  
it is worth discussing as few LBG analogues are 
currently observable in the X-ray without unreasonable
exposure times.  

\begin{figure}
\centering
\columnwidth=.30\columnwidth
\includegraphics[width=3.in]{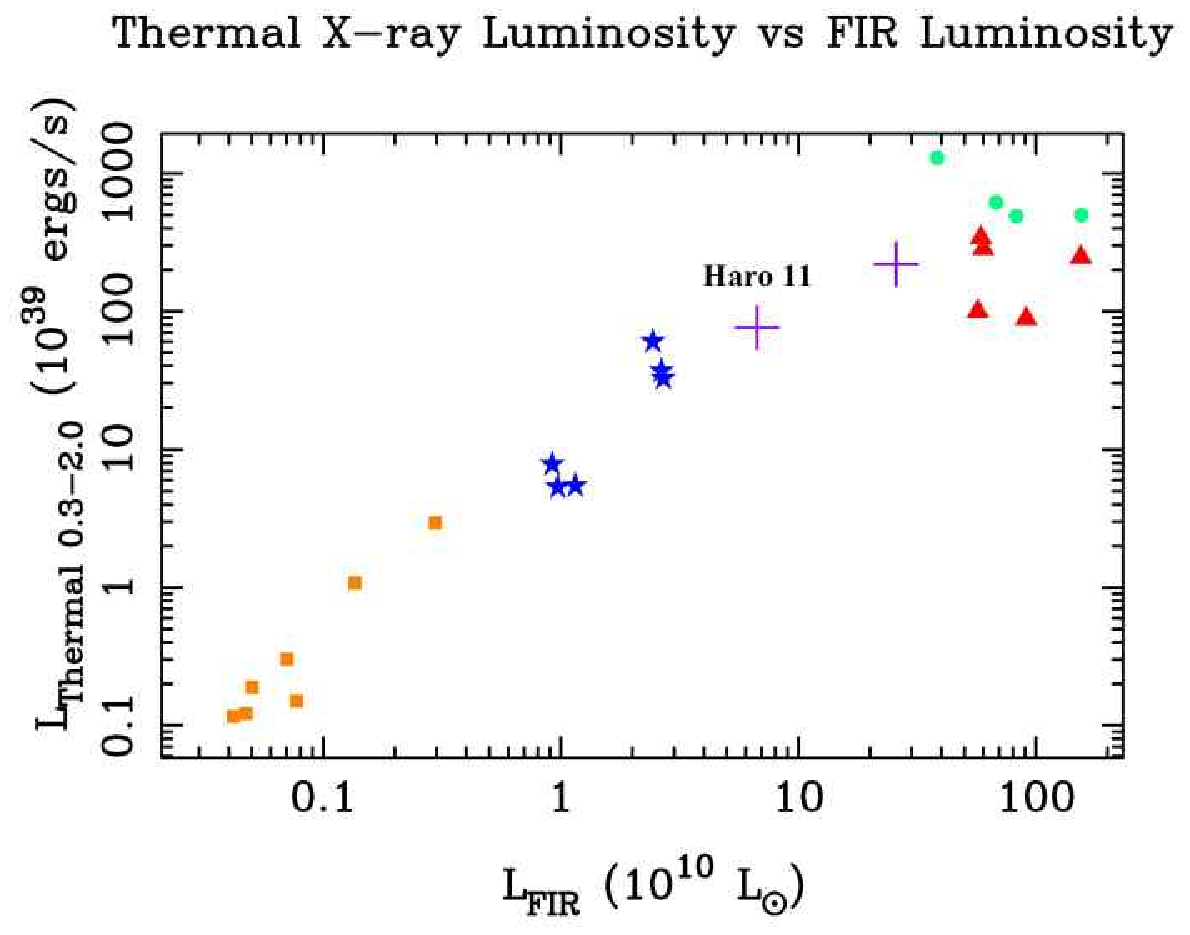}
\includegraphics[width=3.in]{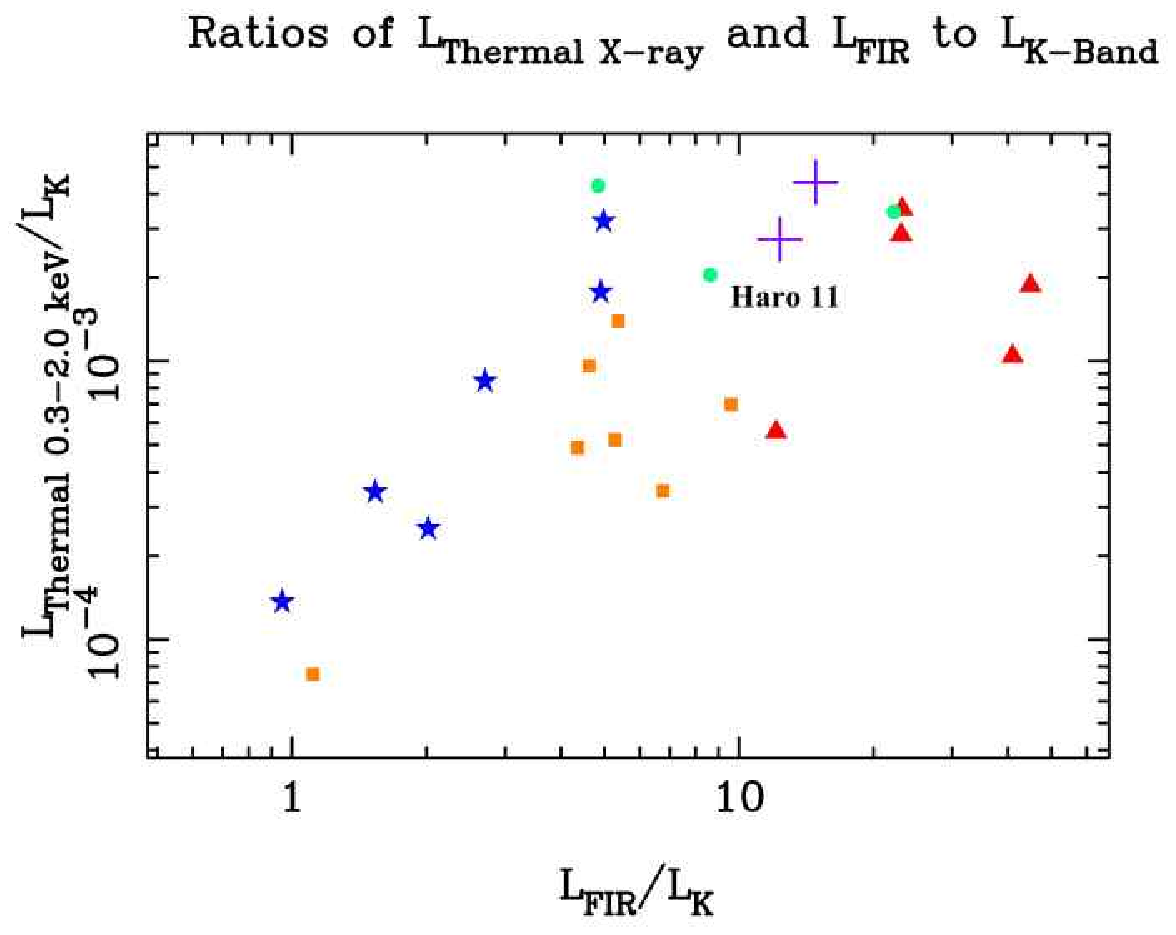}
\vspace{.2in}
\caption{
These plots are originally described in \citet{grim05}.  The left Figure shows a relationship
between $\rm{L_{FIR}\,(SFR)}$ and the thermal X-ray luminosity (0.3-2.0 keV).
On the right we have divided both the thermal X-ray 
and the FIR luminosity by the $K$-band luminosity (a proxy for stellar mass).  The FIR luminosity
of the dwarf starbursts is actually the sum of the their UV and FIR luminosities.
There is a clear
linear relation between the SFR per stellar mass and
the thermal X-ray emission per stellar mass.  The IRAS and $K$-band
data were obtained from {\it NED} and have been tranformed using
the methods of \citet{sand96} and \citet{carp01} respectively.  Both \vv\, and \hoo\,
follow the scaling
relation defined by starbursts.
Symbols indicate dwarf starbursts (orange squares), starbursts (blue stars), \ulg\, (red triangles), AGN \ulg\, (green circles), and local LBG analogues (purple crosses).
\label{div_L_K}}
\end{figure}

\hoo\, and \vv\, were selected due to their high UV luminosities
and compact size.  This suggests a high SFR 
compared to galaxies of similar mass (see \citet{heck05}).  Given the selection criteria,  
we find that the properties of the hot gas follow accordingly.
The left panel of Figure \ref{div_L_K} is a plot of  $\rm{L_{FIR} + L_{UV}}$ 
(a proxy for SFR) 
versus the thermal 0.3-2.0 keV 
X-ray luminosity.  This plot
is originally taken from \citet{grim05} and includes
dwarf starbursts, starbursts, and ULIRGs.
There is a clear correlation (possibly non linear) 
between the thermal X-ray emission and the SFR.
\hoo\, and \vv\, fall between the starbursts 
and ULIRGs as expected for LBG analogues.
In the right-hand panel we have divided the thermal X-ray 
and FIR+UV luminosities by the 
$K$-band luminosity (a proxy for stellar mass).
This shows a fundamental connection between the 
level of star formation and the luminosity of the
thermal X-ray gas.  Unsurprisingly, both \vv\,
and \hoo\, have high SFRs and soft X-ray luminosities given
their stellar mass.

\begin{figure}
\centering
\leavevmode
\columnwidth=.30\columnwidth
\includegraphics[width=3in]{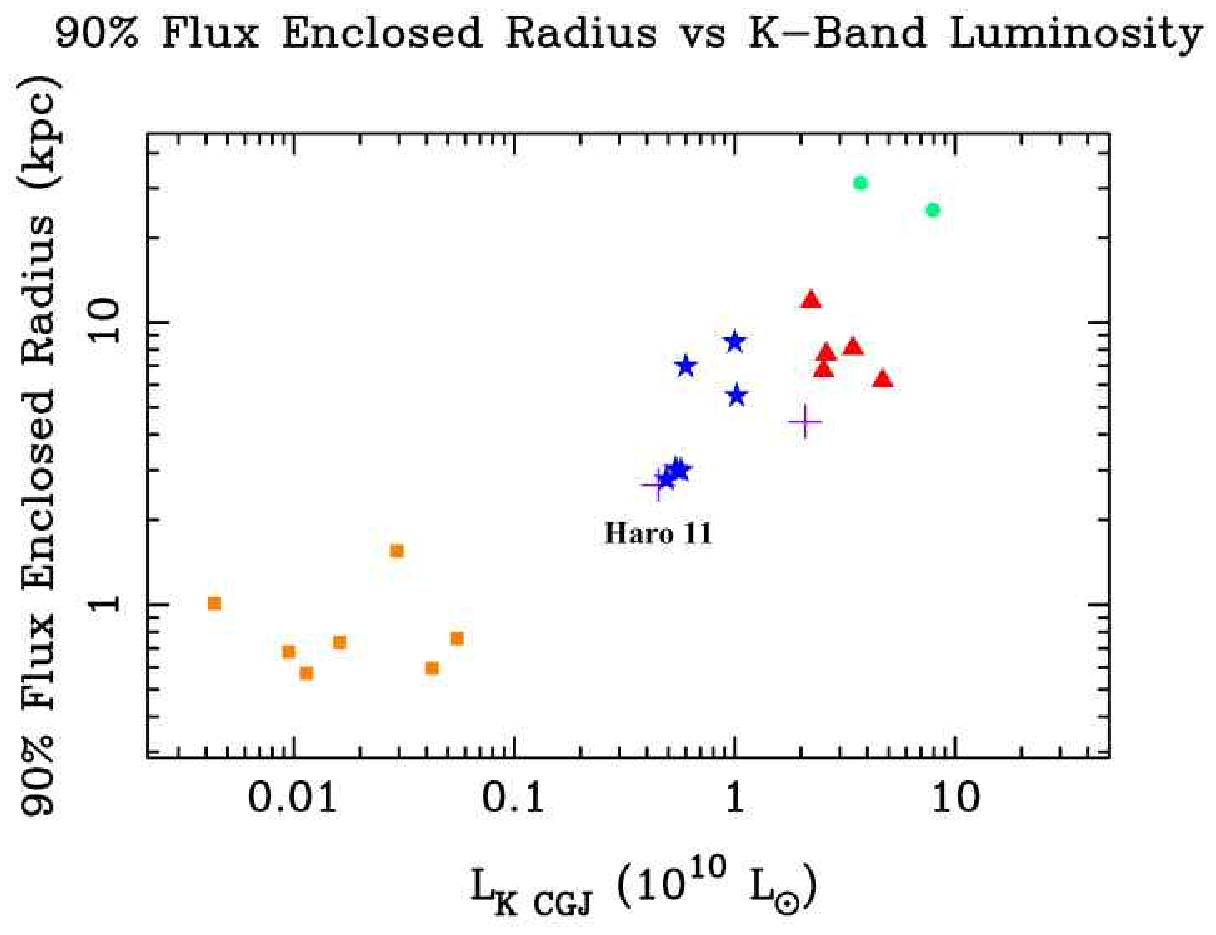}
\hfil
\includegraphics[width=3in]{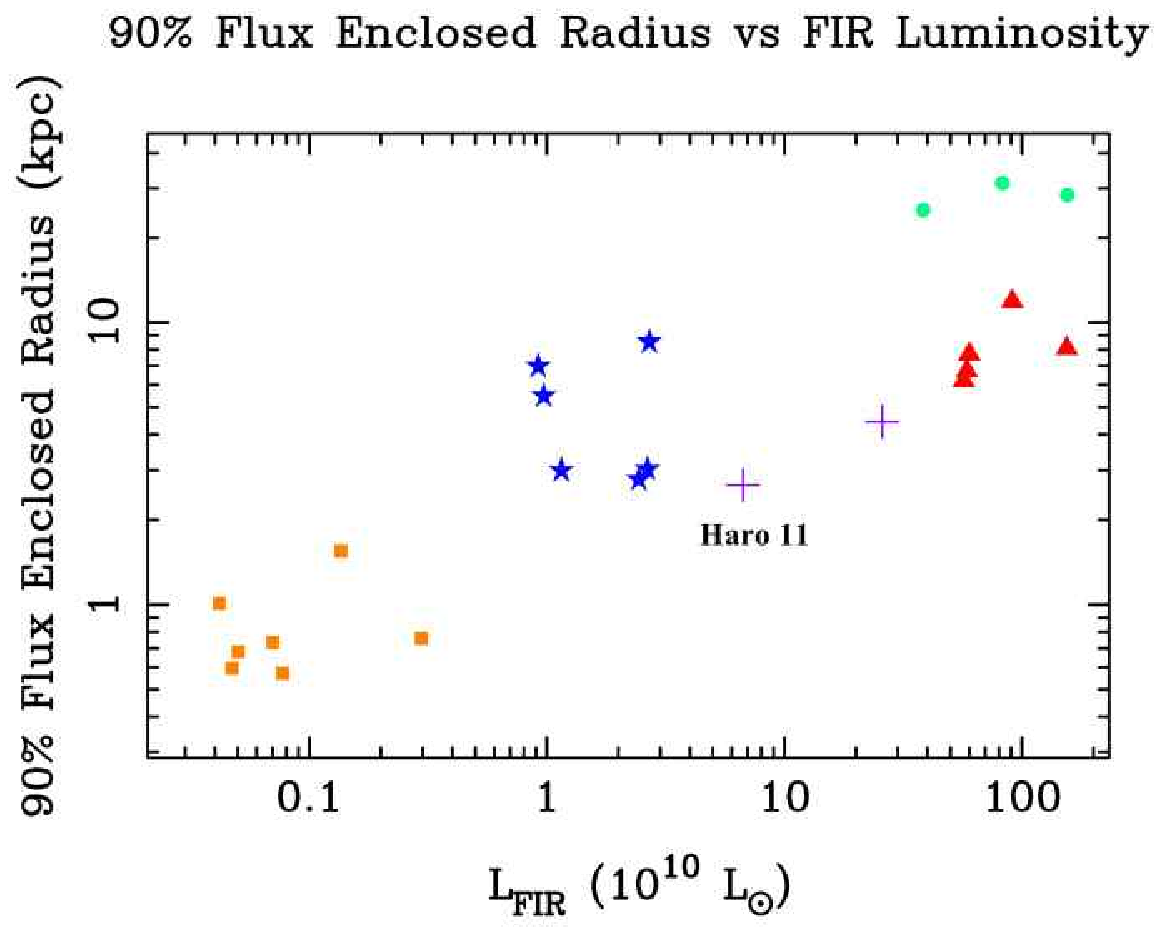}
\caption{
90\% flux enclosed radii  in the 0.3-1.0 keV X-ray band vs $K$-band 
and FIR luminosity (see \citet{grim05} for details).
The size of the X-ray emitting region is correlated with the 
 stellar mass ($K$-band) and SFR (FIR).  The FIR luminosity
of the dwarf starbursts is actually the sum of the their UV and FIR luminosities.
The LBG analogues (\vv\,and\,\hoo)
fall on the low end of FIR relation due to their high SFR and compact size.  
The spatial extent of the hot X-ray gas appears to be more strongly correlated
with the mass of the host galaxy.
Symbols indicate dwarf starbursts (orange squares), 
starbursts (blue stars), \ulg\, (red triangles), AGN \ulg\, (green circles), and 
local LBG analogues (purple crosses).
\label{r90}}
\end{figure}

In \citet{grim05} we showed that both the $K$-band and FIR luminosities
scaled with the radial extent of the soft X-ray emission.  
In Figure \ref{r90} we have plotted the 0.3-1.0 keV 90\% flux
enclosed radius versus the FIR and $K$-band luminosities.
Due to the high SFRs and compact sizes, both
\hoo\, and \vv\, fall below the 
FIR relation for other starbursts.  Previously, the $K$-band
relation had been only marginally preferred
to the FIR-band relation based on
Kendall's rank-order correlation coefficient.
The chances of a spurious correlation
were $2.7\times10^{-5}$ versus
$3.4\times10^{-5}$ for $K$-band and FIR respectively.
With the two new data points 
the chance of a spurious relation is unchanged for the 
FIR data.  However the K-Band relation improved with
a chance of a spurious correlation of 
$4.4\times10^{-6}$.
This suggests that the host galaxy has a strong
effect on the spatial extent of the X-ray emission, 
although the starburst drives
the energetics and composition of the hot gas.

\begin{figure}
\centering
\includegraphics[width=3in]{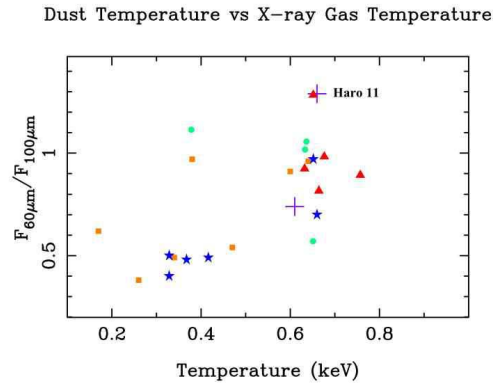}
\vspace{.2in}
\caption{
$\rm{F_{60\mu m}/F_{100\mu m}}$ vs  X-ray gas temperature.
The $\rm{F_{60\mu m}/F_{100\mu m}}$ ratio is an indicator of
dust temperature and hence the SFR per unit area. 
There is a rough correspondence between the 
gas and dust temperatures.
While \vv\, falls near the principal relation, \hoo\,
is an outlier with F60$\micron$/F100$\micron$=1.29.
 Symbols indicate dwarf starbursts (orange squares), starbursts (blue stars), \ulg\, (red triangles), AGN \ulg\, (green circles), and local LBG analogues (purple crosses).
\label{F60o100vkt}}
\end{figure}

Figure \ref{F60o100vkt} has a plot of the X-ray temperature
versus the dust temperature 
(as traced by the IRAS F60$\micron$/F100$\micron$ ratio).
While most of the galaxies show a rough
correlation \hoo\, with F60$\micron$/F100$\micron$=1.29 \citep{soif89} 
is one of the outliers.  Physically, as the 
FIR emission is reprocessed UV light, 
F60$\micron$/F100$\micron$ corresponds to the
luminosity-weighted mean dust temperature
which is set by the mean FUV intensity in the
starburst ( $\propto$ SFR per unit area).
Knot B appears to be a compact region
of intense star formation as it is the brightest feature
in \ha\,, X-ray, Spitzer IRAC and MIPS, and 3.6 cm radio data \citep{schmitt06}. 
Knot B is probably dominating the observed 
IRAS F60$\micron$/F100$\micron$ ratio
and driving the anomalously high SFR per area.


\section{Conclusions}

Star formation driven feedback is an essential ingredient to understanding
galaxy evolution and the IGM.  Observations of local starbursts 
have shown that galactic winds, driven by the kinetic energy from
supernovae and stellar winds, are the strongest manifestation
of star formation driven feedback.  These outflows are complex multi-phase phenomena whose physical, chemical, and dynamical properties can only be understood through complementary observations at many wavebands and detailed modeling \citep{veill05}. Observations of the coronal ($10^5$ to $10^6$ K) and hot ($10^6$ to $10^7$\,K) gas are particularly important, as they provide essential information about the importance of radiative cooling of the outflow and about the dynamics and energy content of the wind. The coronal gas is best traced through the \ion{O}{6} doublet in the far-UV and the hot gas is best observed via its soft X-ray emission.

The strong overall cosmic evolution in the global SFR  \citep[e.g. ][]{bunk04} implies that the bulk of the feedback from galactic winds occurred at early times ($z >$ 1). Indeed, the direct signature of galactic winds -- the presence of broad, blueshifted interstellar absorption lines in the rest-frame UV -- is generically present in the LBGs \citep{shap03}. These are the best-studied population of high-redshift star forming galaxies  \citep{steid99}. However, such data provide only a narrow range of information about galactic winds. Unfortunately, direct observations of the hotter gas in the rest-frame soft X-ray and FUV regions for high redshift galaxies are extremely difficult or even impossible to obtain, which hinders progress in understanding 
this complex phenomenon.  Finding and studying
analogs to LBGs is therefore very important
as a proxy for studying the outflow phenomenon.

In this paper we have described FUV and soft X-ray observations of \hoo, the second Lyman break analogue
observed by \ch.  Our \fu~observations in the FUV show strong and broad interstellar absorption lines. The lines with the highest S/N show two kinematic components. One is centered near the galaxy's  systemic velocity and apparently includes the ISM. The second component is strongly blueshifted,  suggesting an outflowing wind with a velocity of  $\sim200-280\,\ks$.
This is consistent with what is seen at slightly longer rest wavelengths
in high redshift LBGs and further establishes the similarity of \hoo~to the LBGs.  
The \ovi\, feature has a P-Cygni profile that is not seen in other
starbursting galaxies.  While we are not able to clearly determine
whether the source of the \ovi\, emission
is radiatively cooling or resonance scattering, 
we estimate that $\lesssim20\%$ of the available
energy from supernovae could be lost to radiative cooling.  This suggests that 
\ovi\,radiative losses do not significantly inhibit the growth of the outflow in \hoo.

The wind in \hoo ~ might in principle be able to carve a channel in the ISM through
which ionizing photons could escape from the starburst to the IGM. However,
our re-analysis of the \fu\, observations shows that there is no convincing
evidence of Lyman continuum leakage.  This result is in contradiction to a previous
report by \citet{berg06}.

Observations with \ch~of the hot diffuse gas 
in \hoo~are consistent with those seen in a local sample of star forming galaxies with galactic outflows. As expected based on its far-IR luminosity and implied SFR, \hoo ~has X-ray properties intermediate between those of \ulg ~and those of more typical present-day starbursts. It has X-ray properties very similar to that of \vv, the only other LBG analogue mapped 
with \ch.  This suggests
that diffuse thermal X-ray emission should be a common feature of LBGs created in
the shocks between the outflowing wind material and surrounding medium \citep{marc05}.

\acknowledgements
J. G. thanks Henrique Schmitt for providing the \ha\, image of \hoo\,
and Claus Leitherer for helpful discussions on the UV properties of stellar winds.   
Funding for this research was provided
by NASA through Chandra Proposal 8610240.


\clearpage

\begin{deluxetable}{lcccccccccc}
\rotate 
\tabletypesize{\scriptsize}
\tablecolumns{11}
\tablecaption{Comparison of \hoo\, versus \vv\label{prop}}
\tablehead{
\colhead{} & \multicolumn{2}{c}{Position} & \colhead{Distance} & 
\colhead{$\rm{v_{sys}}$} &
\colhead{Scale} & 
\colhead{Z} &
\colhead{$\rm{L_{FIR}}$\tablenotemark{a}} & \colhead{$\rm{L_{K}}$\tablenotemark{a}} &
\colhead{$\rm{L_{FUV}}$\tablenotemark{b}}  &
\colhead{$\rm{I_{FUV}}$\tablenotemark{c}} \\
\colhead{} & \multicolumn{2}{c}{J2000} &  \colhead{Mpc} & \colhead{\kms} & \colhead{kpc arcsec$^{-1}$} & &
\colhead{$\rm{L_\odot}$} & \colhead{$\rm{L_\odot}$} & \colhead{$\rm{L_\odot}$}  &  
\colhead{$\rm{L_\odot\,kpc^{-2}}$} }
\startdata
\hoo & +00 36 52.5 & -33 33 19 & 88 & 6180 & 0.41 & 7.9\tablenotemark{c} & $6.7\times10^{10}$ & $4.5\times10^{9}$ & $2.0\times10^{10}$ & $2.8\times10^{9}$\tablenotemark{d}\\
\vv & +01 07 47.1 & -17 30 24 & 86 & 5970\tablenotemark{e} & 0.40 & 8.6-8.7\tablenotemark{f} & $2.6\times10^{11}$ & $2.0\times10^{10}$ & $2.5\times10^{10}$ 
& $7.8\times10^{8}$\tablenotemark{g}\\
\enddata
\tablenotetext{a}{IRAS \citep{soif89}  and 2MASS \citep{jarr03} results are transformed using methods of \citet{sand96} and \citet{carp01} respectively.}
\tablenotetext{b}{Derived from the \fu\,continuum flux at 1150\,\AA}
\tablenotetext{c}{Gas phase metallicity as log(O/H)+12 from \citet{berg02}.}
\tablenotetext{d}{ACS observation}
\tablenotetext{e}{\citet{grim06}}
\tablenotetext{f}{Gas phase metallicity as log(O/H)+12 from \citet{kim95} using the transformations of \citet{char01}.}
\tablenotetext{g}{GALEX observation}
\end{deluxetable}

\begin{deluxetable}{ccclcccccc}
\rotate
\tabletypesize{\scriptsize}
\tablecolumns{10}
\tablewidth{0pc}
\tablecaption{\hoo\, X-Ray Spectral Fit\tablenotemark{a}}
\tablehead{
\colhead{$\rm{N_H}$\tablenotemark{b}} & \colhead{kT}   & \colhead{$K$\tablenotemark{c}} & \colhead{$\alpha/\rm{Fe}$} & 
\colhead{$N_{H}$} & \colhead{PL Norm\tablenotemark{d}}   
& \colhead{$\Gamma$} & 
\colhead{$\chi^2/\rm{DOF}$} & 
\colhead{$\rm{L_{0.3-2.0\,keV,thermal}}$} & \colhead{$\rm{L_{2.0-8.0\,keV,powerlaw}}$}\\
 \colhead{$10^{22}~\rm{cm^{-2}}$} & \colhead{keV}    & \colhead{} & \colhead{} &
\colhead{$10^{22}~\rm{cm^{-2}}$}   & \colhead{}    & \colhead{} & \colhead{} &
\colhead{\ergs} & \colhead{\ergs}
\label{xspecdata}}
\startdata
$1.9\times10^{-2}$ & $0.68^{+0.05}_{-0.03}$ & $6.0^{+1.4}_{-4.1}\times10^{-5}$ & $4.0^{+1.1}_{-1.2}$ & 
$0.5^{+0.3}_{-0.5}$ & $3.3^{+1.3}_{-1.6}\times10^{-5}$ & $1.7^{+0.3}_{-0.3}$ & 61/65 & $7.2\times10^{40}$ & $1.0\times10^{41}$\\\
\enddata
\tablenotetext{a}{Values derived from an xspec model of  wabs ( vmekal + zwabs ( powerlaw )) using the default abundance settings \citep{and89}.} 
\tablenotetext{b}{Galactic $\rm{N_H}$ value fixed to COLDEN result}
\tablenotetext{c}{Plasma model normalization in units of 
$\frac{10^{-14}}{4\pi [D_A (1+z)]^2}\int n_e n_H dV$, where $D_A$ is the angular distance, and $n_H$ and $n_e$ are the hydrogen and electron number densities respectively.}
\tablenotetext{d}{$\rm{photons~keV^{-1}\,cm^{-2}\, s^{-1}}$ at 1 keV}
\end{deluxetable}

%

\clearpage

\begin{deluxetable}{lcccccc}
\tabletypesize{\footnotesize}
\tablecolumns{7}
\tablewidth{0pc}
\tablecaption{ISM Absorption Line Fit Data}
\tablehead{
\colhead{Ion} & \colhead{$\lambda_{0}$} & \colhead{$log(\lambda f N/N_H)$} &
                 \colhead{Instrument} & \colhead{$\rm{W}_\lambda$} & 
                 \colhead{$v_{c}$} & \colhead{FWHM} \\
                  & \colhead{$\ang$} & & & \colhead{$\ang$} & \colhead{\kms} & 
                  \colhead{\kms}\label{absdata}}
\startdata
Ly~$\beta$\tablenotemark{b} & 1025.722 & 1.91 & LiF 1A &  $2.3\pm0.1$ & $6055\pm9$ & $557\pm31$\\
 & & & LiF 2B &  $2.3\pm0.1$ & $6051\pm9$ & $529\pm54$\\
\ion{C}{3}\tablenotemark{b} & 977.02 & -0.61 & LiF 1A &  $0.9\pm0.2$ & 6081\tablenotemark{a} & $169\pm18$\\
 & & & LiF 2B &  $0.7\pm1.4$ & $6133$\tablenotemark{a} & $87\pm14$\\
 \ovi & 1031.926 & -1.13 & LiF 1A &  $0.4\pm0.1$ & $6042\pm16$ & $193\pm35$\\
 & & & LiF 2B &  $0.4\pm0.1$ & $6051\pm15$ & $159\pm34$\\
 \ion{C}{2} & 1036.337 & -1.37 & LiF 1A &  $1.4\pm0.1$ & $6061\pm6$ & $402\pm19$\\
 & & & LiF 2B &  $1.2\pm0.1$ & $6075\pm8$ & $341\pm30$\\
\ion{O}{1} & 988.733 & -1.56 & LiF 1A &  $0.7\pm0.1$ & $6092\pm13$ & $275\pm15$\\
 & & & LiF 2B &  $0.7\pm0.1$ & $6079\pm23$ & $304\pm26$\\
\ion{N}{2} & 1083.99 & -2.00 & LiF 1B &  $0.9\pm0.1$ & $6081\pm10$ & $352\pm28$\\
 & & & LiF 2A &  $1.0\pm0.1$ & $6085\pm11$ & $381\pm33$\\
\ion{N}{3} & 989.799 & -2.03 &  LiF 1A & $1.0\pm0.1$ & $6076\pm6$ & $275\pm137$\\
 & & & LiF 2B &  $1.1\pm0.1$ & $6073\pm10$ & $304\pm152$\\
\ion{N}{1} & 1134.415 & -2.09 & LiF 2A &  $0.2\pm0.1$ & $6074$\tablenotemark{a} & $103\pm36$\\
\ion{O}{1} & 1039.23 & -2.29 & LiF 1A &   $0.3\pm0.1$ & $6110\pm5$ & $207\pm47$\\
 & & & LiF 2B &  $0.3\pm0.1$ & $6135\pm13$ & $192\pm46$\\
\ion{Fe}{2} & 1096.877 & -2.96 & LiF 1B &  $0.3\pm0.1$ & $6101\pm15$ & $197\pm72$\\
 & & & LiF 2A &  $0.2\pm0.1$ & $6127\pm17$ & $143\pm40$\\
 \ion{Si}{2} & 1020.699 & -2.989 & LiF 1A &  $0.1\pm0.1$ & $6071\pm37$ & $87\pm83$\\
 & & & LiF 2B &  $0.2\pm0.1$ & $6071\pm12$ & $100\pm23$\\
 \ion{Fe}{2} & 1121.975 & -3.15 & LiF 2A &  $0.3\pm0.2$ & $6088\pm68$ & $203\pm193$\\
\ion{S}{3} & 1012.502 & -3.24 & LiF 1A &  $0.5\pm0.2$ & $6083\pm6$ & $225\pm108$\\
 & & & LiF 2B &  $0.6\pm0.2$ & $6078\pm13$ & $290\pm75$\\
\ion{Fe}{2} & 1125.448 & -3.41 & LiF 2A &  $0.2\pm0.1$ & $6062\pm17$ & $195\pm37$\\
\ion{Ar}{1} & 1066.66 & -3.59 & LiF 2A &  $0.5\pm0.2$ & $6136\pm27$ & $332\pm82$\\
\enddata
\tablenotetext{a}{Errors are undetermined for this value.}
\tablenotetext{b}{$\rm{Ly}$~$\beta$ and \ion{C}{3} are saturated at line center.}
\end{deluxetable}

\clearpage
\begin{deluxetable}{lccccc}
\tabletypesize{\footnotesize}
\tablecolumns{6}
\tablewidth{0pc}
\tablecaption{Stellar Photospheric Absorption Line Fit Data}
\tablehead{
\colhead{Ion} & \colhead{$\lambda_{0}$} &
                 \colhead{Instrument} & \colhead{$\rm{W}_\lambda$} & 
                 \colhead{$v_{c}$} & \colhead{FWHM} \\
                  & \colhead{$\ang$} & & \colhead{$\ang$} & \colhead{\kms} & 
                  \colhead{\kms}\label{phabsdata}}
\startdata
\ion{Si}{4} & 1122.487 & LiF 2A & $0.6\pm0.1$ & $6163\pm17$ & $222\pm29$\\
\ion{Si}{4} & 1128.201 & LiF 2A & $0.3$\tablenotemark{a} & $6202$\tablenotemark{a} & $327$\tablenotemark{a}\\
\ion{P}{5} & 1117.977 & LiF 2A & $0.3\pm0.1$ & $6165\pm24$ & $320\pm63$\tablenotemark{a}\\
\ion{P}{5} & 1128.0 & LiF 2A & $0.1$\tablenotemark{a} & $6202$\tablenotemark{b} & $327$\tablenotemark{b}\\
\enddata
\tablenotetext{a}{Errors are undetermined for this value.}
\tablenotetext{b}{Values of the velocity shift and FWHM of \ion{P}{5} $\lambda$1128 are tied to 
the corresponding parameters in \ion{Si}{4} $\lambda$1128}
\end{deluxetable}

\clearpage
\begin{deluxetable}{lccccccccccc}
\rotate 
\tabletypesize{\scriptsize}
\tablecolumns{12}
\tablewidth{0pc}
\tablecaption{Two Component Absorption Line Fits}
\tablehead{
& & & & & \multicolumn{3}{c}{Galaxy} & & \multicolumn{3}{c}{Outflow} \\
\cline{6-8} \cline{10-12}
\colhead{Ion} & \colhead{$\lambda_{0}$} & \colhead{$log(\lambda f N/N_H)$} &
                 \colhead{Instrument} & \phantom{ } & \colhead{$\rm{W}_\lambda$} & 
                 \colhead{$v_{c}$} & \colhead{FWHM} & \phantom{ } & \colhead{$\rm{W}_\lambda$} & 
                 \colhead{$v_{c}$} & \colhead{FWHM} \\
                  & \colhead{$\rm{\AA}$} & & & & \colhead{$\rm{\AA}$} & \colhead{\kms} & 
                  \colhead{\kms} & & \colhead{$\rm{\AA}$} & \colhead{\kms} & 
                  \colhead{\kms}\label{twoabsdata}}
\startdata
Ly~$\beta$\tablenotemark{a} & 1025.722 & 1.91 & LiF 1A &  & $2.22\pm0.07$ & $6060\pm7$ & $519\pm19$ & & $0.25\pm0.09$ & $5781\pm18$ & $124\pm40$\\
 & & & LiF 2B & &$1.91\pm0.15$ & $6103\pm20$ & $467\pm31$ & & $0.56\pm0.16$ & $5827\pm21$ & $184\pm47$\\ 
\ion{C}{3}\tablenotemark{a}  & 977.02 & -0.61 & LiF 1A &  & $0.85\pm0.18$ & $6084\pm6$ & $146\pm13$ & & $0.44\pm0.18$ & $5840\pm7$ & $60$\tablenotemark{b} \\
 & & & LiF 2B & &$0.69$\tablenotemark{b} & $6133\pm4$ & $83\pm9$ & & $0.36$\tablenotemark{b}  & $5894\pm6$ & $53\pm25$\\ 
\ion{C}{2} & 1036.337 & -1.374 & LiF 1A &  & $1.22\pm0.09$ & $6076\pm8$ & $356\pm28$ & & $0.11\pm0.05$ & $5855\pm27$ & $100$\tablenotemark{b} \\
 & & & LiF 2B & &$1.18\pm0.10$ & $6084\pm12$ & $325\pm26$ & & $0.10\pm0.17$ & $5866\pm70$ & $104\pm164$\\ 
\ion{N}{2} & 1083.99 & -2.002 & LiF 1B &  & $0.79\pm0.14$ & $6104\pm12$ & $310\pm50$ & & $0.13\pm0.05$ & $5935\pm11$ & $88\pm26$\\
 & & & LiF 2A & & $1.05\pm0.20$ & $6114\pm20$ & $392\pm63$ & & $0.07\pm0.20$ & $5921\pm39$ & $123$\tablenotemark{b} \\
\ion{N}{3}\tablenotemark{c} & 989.799 & -2.03 & LiF 1B &  & $0.81\pm0.04$ & $6099\pm11$ & $217\pm27$ & & $0.35\pm0.13$ & $5957\pm22$ & $150\pm40$\\
 & & & LiF 2A & & $0.78\pm0.22$ & $6120\pm31$ & $268\pm41$ & & $0.41\pm0.17$ & $6022\pm17$ & $128\pm38$\\
\enddata
\tablenotetext{a}{The galaxy and/or outflow component of this line is saturated.}
\tablenotetext{b}{Unable to determine errors on this value.}
\tablenotetext{c}{\ion{N}{3}~$\lambda$990 is blended with \ion{O}{1}~$\lambda$989.  
They are fit simultaneously with the tied values of the fwhm and relative velocity.}
\end{deluxetable}

\begin{deluxetable}{lcccccc}
\tabletypesize{\footnotesize}
\tablecolumns{7}
\tablewidth{0pc}
\tablecaption{Mass and Energy Outflow Rates}
\tablehead{
\colhead{Ion}  &  \colhead{$\rm{N_{Ion}}$\tablenotemark{a}} & 
\colhead{Abundance\tablenotemark{b}} & \colhead{$\rm{N_H}$}   & 
\colhead{$\frac{\Omega}{4\pi}\,\dot{\rm{M}}$\tablenotemark{c}} & 
\colhead{$\frac{\Omega}{4\pi}\dot{\rm{E}}$\tablenotemark{c}}\\
\colhead{} & \colhead{$10^{14}~\rm{cm^{-2}}$}   & \colhead{$\rm{\log(N_{Elem}/N_H)+12}$}    & \colhead{$10^{18}~\rm{cm^{-2}}$} & 
\colhead{$\rm{M_\odot/yr}$} & \colhead{$\rm{10^{40}~erg/s}$}\label{outflow}}
\startdata
\ion{N}{2} & 1.4 & 7.78 & 2.3 & 0.1 & 0.2 \\
\ion{N}{3} & 5.7 & 7.78 & 9.5 & 0.3 & 0.7 \\
\ion{C}{2} & 3.6 & 8.39 & 1.5 & 0.1 & 0.1\\
Total Cool Gas\tablenotemark{d}     & & & $>$11.8 & $>$0.4 & $>$0.9\\
\cline{1-6} 
Total Hot Gas\tablenotemark{e}       & & &  & $18f^{1/2}$ & $130f^{1/2}$\\ 
\enddata
\tablenotetext{a}{Based on outflow equivalent width and FWHM from Table \ref{twoabsdata}.}
\tablenotetext{b}{\citet{asp05}}
\tablenotetext{c}{Estimated mass and kinetic energy outflow rates (see text).} 
\tablenotetext{d}{From \fu\, data using \ion{N}{2} and \ion{N}{3}. } 
\tablenotetext{e}{From X-ray data.  Note that f is the volume filling factor of the hot gas.} 
\end{deluxetable}

\clearpage

\end{document}